\documentclass[]{mn2e}
\usepackage{epsfig,times}

\newcommand{\etal}{{et~al.\ }}  

\title
{A binary model for the UV-upturn of elliptical galaxies}

\author[Han, Podsiadlowski \mbox{\rm\&} Lynas-Gray]
{Z.~Han$^1$\thanks{E-mail: zhanwenhan@hotmail.com}, Ph.~Podsiadlowski$^2$,
A.E. Lynas-Gray$^2$\\
\it
$^1$ National Astronomical Observatories / Yunnan Observatory,
the Chinese Academy of Sciences, Kunming, 650011, China\\
$^2$ University of Oxford, Department of Physics, Keble Road, Oxford,
OX1 3RH
}
\begin{document}
\maketitle

\begin{abstract}
The discovery of a flux excess in the far-ultraviolet (UV)
spectrum of elliptical galaxies was a major surprise in 1969.  While
it is now clear that this UV excess is caused by an old population of
hot helium-burning stars without large hydrogen-rich envelopes, rather
than young stars, their origin has remained a mystery.  Here we show
that these stars most likely lost their envelopes because of binary
interactions, similar to the hot subdwarf population in our own
Galaxy. We have developed an evolutionary population synthesis model
for the far-UV excess of elliptical galaxies based on the binary
model developed by Han \etal (2002, 2003) for the formation of hot
subdwarfs in our Galaxy.  Despite its simplicity, it successfully
reproduces most of the properties of elliptical galaxies with a UV
excess: the range of observed UV excesses, both in $(1550-V)$
and $(2000-V)$, and their evolution with redshift.  We also present
colour-colour diagrams for use as diagnostic tools in
the study of elliptical galaxies.  The model has major implications
for understanding the evolution of the UV excess and of elliptical
galaxies in general. In particular, it implies that the UV excess is
not a sign of age, as had been postulated previously, and predicts
that it should not be strongly dependent on the metallicity of the
population, but exists universally from dwarf ellipticals to giant
ellipticals.
\end{abstract}


\section{Introduction}

A long-standing problem in the study of elliptical galaxies is the
far-ultraviolet (UV) excess in their spectra. Traditionally,
elliptical galaxies were supposed to be passively evolving and not 
contain any young stars that radiate in the far-UV.  Therefore, the
discovery of an excess of radiation in the far-UV by the {\it Orbiting
Astronomical Observatory} mission 2 (OAO-2)
\cite{cod69} came as a complete surprise. Indeed, this was one of the
first major discoveries in UV astronomy and became a basic property of
elliptical galaxies.  This far-UV excess is often referred to as the
``UV-upturn'', since the flux increases in the spectral energy
distributions of elliptical galaxies as the wavelength decreases from
2000 to 1200\AA. The UV-upturn is also known as UV rising-branch, UV
rising flux, or simply UVX (see the review by O'Connell 1999).

The UV-upturn phenomenon exists in virtually all elliptical galaxies
and is the most variable photometric feature.  A famous correlation
between UV-upturn magnitude and metallicity was found by 
Burstein \etal 
\shortcite{bur88} from {\it International Ultraviolet Explorer
Satellite} (IUE) spectra of 24 quiescent early-type galaxies
(hereafter BBBFL relation). The UV-upturn could be important in many
respects: for the formation history and evolutionary properties
of stars, the chemical enrichment of galaxies, galaxy dynamics,
constraints to the stellar age and metallicity of galaxies and
realistic ``K-corrections'' \cite{oco99,yi98,bro04,yi04}. In
particular, the UV-upturn has been proposed as a possible age
indicator for giant elliptical galaxies \cite{bre96,chi97,yi99}.  The
origin of the UV-upturn, however, has remained one of the great
mysteries of extragalactic astrophysics for some 30 years
\cite{bro04}, and numerous speculations 
have been put forward to explain it:
non-thermal radiation from an active galactic nucleus (AGN), young massive
stars, the central stars of planetary nebulae (PNe) or post-asymptotic
giant branch (PAGB) stars, horizontal branch (HB) stars and post-HB
stars (including post-early AGB stars and AGB-manqu\'e stars) and
accreting white dwarfs \cite{cod69,hil71,gun81,nes85,moc86,kja87,gre90}.
Using observations made with the 
the Hopkins Ultraviolet Telescope (HUT) and comparing them to synthetic
spectra, Ferguson \etal \shortcite{fer91} and subsequent studies
by Brown, Ferguson \& Davidsen
\shortcite{bro95}, Brown \etal \shortcite{bro97}, and Dorman,
O'Connell \& Rood \shortcite{dor95} were able to show 
that the UV upturn is mainly caused by extreme
horizontal branch (EHB) stars.  Brown \etal \shortcite{bro00b}
detected EHB stars for the first time in an elliptical galaxy (the
core of M~32) and therefore provided direct evidence for the EHB
origin of the UV-upturn.

EHB stars, also known as subdwarf B (sdB) stars, are
core-helium-burning stars with extremely thin hydrogen envelopes
($M_{\rm env}\le 0.02M_\odot$), and most of them are believed to have
masses around $0.5M_\odot$ \cite{heb86,saf94}, as has recently been
confirmed asteroseismologically in the case of PG 0014+067
\cite{bra01}.  They have a typical luminosity of a few $L_\odot$, a
lifetime of $\sim 2\times 10^8 {\rm yr}$, and a characteristic surface
temperature of $\sim 25\ 000{\rm K}$ \cite{dor93,dcr96,han02}. The
origin of those hot, blue stars, as the major source of the far UV
radiation, has remained an enigma in evolutionary population synthesis
(EPS) studies of elliptical galaxies.  Two models, both involving
single-star evolution, have previously been
proposed to explain the UV-upturn: a metal-poor model
\cite{lee94,par97} and a metal-rich model
\cite{bre94,bre96,tan96,yi95,yi97a,yi97b,yi98}.  

The metal-poor model ascribes the UV-upturn to an old metal-poor
population of hot subdwarfs, blue core-helium-burning stars, that
originate from the low-metallicity tail of a stellar population with a
wide metallicity distribution \cite{lee94,par97}.  This model explains
the BBBFL relation since elliptical galaxies with high average
metallicity tend to be older and therefore have stronger UV-upturns.
The model tends to require a very large age of the population (larger
than the generally accepted age of the Universe), and it is not clear
whether the metal-poor population is sufficiently blue or
not. Moreover, the required low-metallicity appears inconsistent with
the large metallicity inferred for the majority of stars in elliptical
galaxies \cite{zho92,ter02,tho05}.

In the metal-rich model the UV-upturn is caused by metal-rich stars
that lose their envelopes near the tip of the first-giant branch
(FGB). This model \cite{bre94,yi97b} assumes a relatively high
metallicity -- consistent with the metallicity of typical elliptical
galaxies ($\sim 1-3$ times the solar metallicity).  In the model, the
mass-loss rate on the red-giant branch, usually scaled with the
Reimer's rate \cite{rei75}, is assumed to be enhanced, where the
coefficient $\eta_{\rm R}$ in the Reimer's rate is taken to be $\sim
2-3$ times the canonical value. In order to reproduce the HB morphology of
Galactic globular clusters, either a broad distribution of
$\eta_{\rm R}$ is postulated \cite{dcr96}, or, alternatively, 
a Gaussian mass-loss 
distribution is applied that is designed to reproduce
the distribution of horizontal-branch stars of a given age and metallicity
\cite{yi97b}.  This model also needs a population age that
is generally larger than 10\,Gyr.  To explain the BBBFL UV-upturn --
metallicity relation, the Reimer's coefficient $\eta_{\rm R}$ is assumed
to increase with metallicity, and the enrichment parameter for the
helium abundance associated with the increase in metallicity, ${\Delta
Y \over \Delta Z}$, needs to be $> 2.5$.

Both of these models are quite {\it ad hoc}: there is neither
observational evidence for a very old, low-metallicity sub-population
in elliptical galaxies, nor is there a physical explanation for the
very high mass loss required for just a small subset of
stars. Furthermore, the onset of the formation of the hot subdwarfs
is very sudden as the stellar population evolves, and both models require a
large age for the production of the hot stars.  As a consequence, the
models predict that the UV upturn of elliptical galaxies declines
rapidly with redshift. However, this does not appear to be consistent
with recent observations with the {\it Hubble Space Telescope} (HST)
\cite{bro98,bro00a,bro03}. The recent survey with the {\it Galaxy
Evolution Explorer} (GALEX) satellite \cite{ric05} showed that the
intrinsic UV-upturn seems not to decrease in strength with redshift.

The BBBFL relation shows that the $(1550-V)$ colour becomes bluer with
metallicity (or Lick spectral index Mg$_{\rm 2}$), where $(1550-V)$ is
the far-UV magnitude relative to the $V$ magnitude.  The relation
could support the metal-rich model.  However, the correlation is far
from being conclusively established.  Ohl \etal \shortcite{ohl98}
studied the far-UV colour gradients in 8 early-type galaxies and found
no correlation between the FUV-{\it B} colour gradients and the
internal metallicity gradients based on the Mg$_{\rm 2}$ spectral line
index, a result not expected from the BBBFL relation.  Deharveng,
Boselli \& Donas \shortcite{deh02} studied the far-UV radiation of 82
early-type galaxies, a UV-flux selected sample, and compared them to
the BBBFL sample, investigating individual objects with a substantial
record in the refereed literature spectroscopically\footnote{Note,
however, that some of the galaxies show hints of recent star formation
\cite{deh02}.}.  They found no correlation between the $(2000-V)$
colour and the Mg$_{\rm 2}$ spectral index. Rich \etal
\shortcite{ric05} also found no correlation in a sample of 172 red
quiescent early-type galaxies observed by GALEX and the {\it Sloan
Digital Sky Survey} (SDSS). Indeed, if there is a weak correlation in
the data, the correlation is in the opposite sense to that of BBBFL:
metal-rich galaxies have redder $(FUV-r)_{\rm AB}$ (far-UV
magnitude minus red magnitude).  On the other hand, Boselli \etal
\shortcite{bos05}, using new GALEX data, reported a mild positive correlation
between $(FUV-NUV)_{\rm AB}$, which is the far-UV magnitude relative
to the near-UV, and metallicity in a sample of early-type galaxies in
the Virgo Cluster.  Donas \etal \shortcite{don06} use GALEX photometry
to construct colour-colour relationships for nearby early-type
galaxies sorted by morphological type. They also found a marginal
positive correlation between $(FUV-NUV)_{\rm AB}$ and
metallicity. These correlations, however, do not necessarily support
the BBBFL relation, as neither Boselli \etal \shortcite{bos05} nor
Donas \etal \shortcite{don06} show that $(FUV-r)_{\rm AB}$ correlates
significantly with metallicity.  Therefore, this apparent lack of an 
observed correlation between the strength of the UV-upturn and metallicity
casts some doubt on the metal-rich scenario.

Both models ignore the effects of binary evolution.  On the
other hand, hot subdwarfs have long been studied in our own Galaxy
\cite{heb86,gre86,dow86,saf94}, and it is now well established that
the vast majority of (and quite possibly all) Galactic hot subdwarfs
are the results of binary interactions.  Observationally, more than
half of Galactic hot subdwarfs are found in binaries
\cite{fer84,all94,the95,ull98,azn01,max01,wil01,ree04}, and orbital
parameters have been determined for a significant sample
\cite{jef98,koe98,saf98,kil99,mor99,oro99,woo99,max00a,max00b,max01,nap01,heb02,heb04,mor04,cha05,mor06}. There has also been substantial theoretical
progress \cite{men76,web84,ibe86,tut90,dcr96,swe97}.  Recently, Han \etal
\shortcite{han02,han03} proposed a binary model (hereafter HPMM model)
for the formation of hot subdwarfs in binaries and single hot
subdwarfs.  In the model, there are three formation channels for hot
subdwarfs, involving common-envelope \cite{pac76} ejection for hot
subdwarf binaries with short orbital periods, stable Roche lobe
overflow for hot subdwarfs with long orbital periods, and the merger
of helium white dwarfs to form single hot subdwarfs.  The model can
explain the main observational characteristics of hot subdwarfs, in
particular, their distributions in the orbital period--minimum
companion mass diagram and in the effective temperature--surface
gravity diagram, their distributions of orbital period and mass
function, their binary fraction and the fraction of hot subdwarf
binaries with white dwarf (WD) companions, their birth rates and their
space density.  More recent observations (e.g.\ Lisker \etal 2004,
2005) support the HPMM model, and the model is now widely used in the
study of hot subdwarfs (e.g.\ Heber \etal 2004, Morales-Rueda, Maxted
\& Marsh 2004, Charpinet \etal 2005, Morales-Rueda \etal 2006).

Hot subdwarfs radiate in UV, and we can apply the HPMM scenario
without any further assumptions to the UV-upturn problem of elliptical
galaxies. The only assumption we have to make is that the stellar
population in elliptical galaxies, specifically their binary properties,
are similar to those in our own galaxy. Indeed, as we will show in
this paper, the UV flux from hot subdwarfs produced in binaries is
important. This implies that any model for elliptical galaxies that
does not include these binary channels is {\em necessarily} incomplete
or has to rely on the {\em apriori} implausible assumption that the
binary population in elliptical galaxies is intrinsically
different. The main purpose of this paper is to develop an {\em
apriori} EPS model for the UV-upturn of elliptical galaxies, by
employing the HPMM scenario for the formation of hot subdwarfs, and to
compare the model results with observations.

The outline of the paper is as follows.
We describe the EPS model in Section 2 and the simulations in Section 3.
In Section 4 we present the results and discuss them, and end the
paper with a summary and conclusions in Section 5.

\section{The Model}

EPS is a technique for modelling the
spectrophotometric properties of a stellar population using our
knowledge of stellar evolution.  The technique was first devised by
Tinsley \cite{tin68} and has experienced rapid
development ever since
\cite{bru93,wor94,bre94,tan96,zha02,bru03,zha04a}.  Recently, binary
interactions have also been incorporated into EPS studies
\cite{zha04b,zha05a,zha05b,zha06} with the rapid binary-evolution code
developed by Hurley, Tout \& Pols \shortcite{hur02}.  In the present paper
we incorporate the HPMM model into EPS by adopting the
binary population synthesis (BPS) code of 
Han \etal \shortcite{han03}, 
which was designed to investigate the formation of many interesting
binary-related objects, including hot subdwarfs.

\subsection{The BPS code and the formation of hot subdwarfs}
\begin{figure*}
\vskip 2cm
\epsfig{file=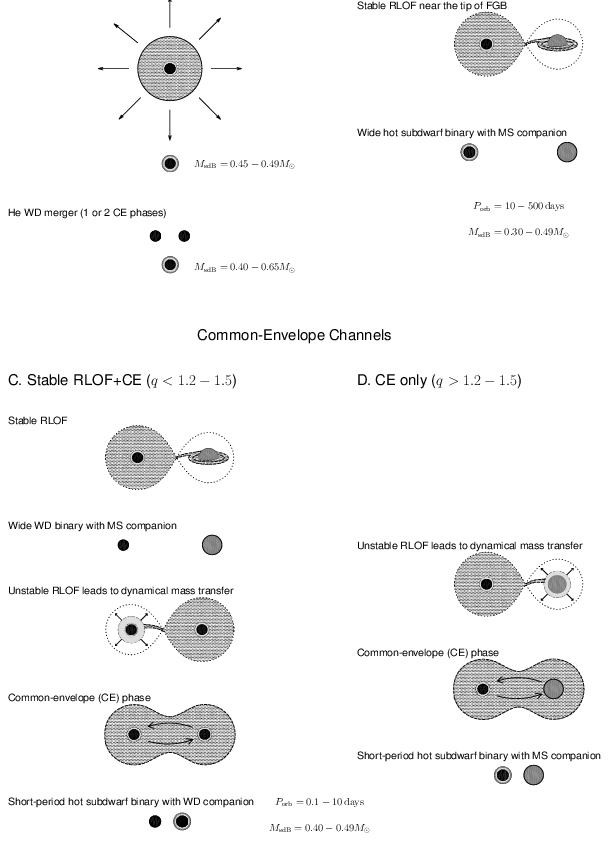,width=14cm}
\caption{
Single and binary channels to produce hot
subdwarfs, core-helium-burning stars with
no or small hydrogen-rich envelopes. {\bf (A)} Single hot subdwarfs may
result from large mass loss near the tip of the first giant branch (FGB),
as in the metal-rich model, or from the merger of two helium
white dwarfs. {\bf (B)} Stable Roche lobe overflow (RLOF) near
the tip of the FGB produces hot subdwarfs in wide binaries.
{\bf (C + D)} Common-envelope evolution leads to hot subdwarfs in very
close binaries, where the companion can either be a white dwarf
(C) or a main-sequence star (D). The simulations presented in this paper
include all channels except for the metal-rich single star channel.
}
\label{chart}
\end{figure*}

The BPS code of Han \etal
was originally developed in 1994 and has been updated regularly ever since
\cite{han94,han95a,han95b,han95c,han98,han02,han03,han04}.
With the code, millions of stars
(including binaries) can be evolved simultaneously from the 
zero-age main sequence (ZAMS) 
to the white-dwarf (WD) stage or a supernova explosion.
The code can simulate in a Monte-Carlo way the formation of
many types of stellar objects,
such as Type Ia supernovae (SNe Ia), double degenerates (DDs),
cataclysmic variables (CVs), barium stars, planetary nebulae (PNe)
and hot subdwarfs.
Note that ``hot subdwarfs'' in this paper is used as a collective term
for subdwarf B stars, subdwarf O stars, and subdwarf OB stars.
They are core-helium-burning stars with thin hydrogen envelopes
and radiate mainly in the UV
(see Figure~\ref{chart} for the formation
channels of hot subdwarfs).

The main input into the BPS code is a grid of stellar models.  For the
purpose of this paper, we use a Population I (pop I) grid with a
typical metallicity $Z=0.02$. The grid, calculated with Eggleton's
stellar evolution code \cite{egg71,egg72,egg73,han94,pol95,pol98},
covers the evolution of normal stars in the range of
$0.08-126.0M_\odot$ with hydrogen abundance $X=0.70$ and helium
abundance $Y=0.28$, helium stars in the range of $0.32-8.0M_\odot$ and
hot subdwarfs in the range of $0.35-0.75M_\odot$ (see Han \etal
2002, 2003 for details).  Single stars are evolved via interpolations
in the model grid.  In this paper, we use $t_{\rm FGB}$ instead of
$\log m$ as the interpolation variable between stellar evolution
tracks, where $t_{\rm FGB}$ is the time from the ZAMS to the tip of
the FGB
for a given stellar mass $m$ and is calculated from fitting
formulae. This is to avoid artefacts in following the time evolution
of hot subdwarfs produced from a stellar population.

The code needs to model the evolution of binary stars as well as of
single stars.  The evolution of binaries is more complicated due to
the occurrence of Roche lobe overflow (RLOF).  The binaries of 
main interest here usually experience two phases of RLOF: the first
when the primary fills its Roche lobe which may produce a WD binary,
and the second when the secondary fills its Roche lobe.

The mass gainer in the first RLOF phase is most likely a 
main-sequence (MS)
star. If the mass ratio $q=M_1/M_2$ at the onset of RLOF is lower than a
critical value, $q_{\rm crit}$, RLOF is stable
\cite{pac65,pac69,pla73,hje87,web88,sob97,han01}. For systems
experiencing their first phase of RLOF in the Hertzsprung gap, we use
$q_{\rm crit}=3.2$ as is supported by detailed binary-evolution
calculations of Han \etal \shortcite{han00}.
For the first RLOF phase on the FGB or 
asymptotic giant branch (AGB),
we mainly use $q_{\rm crit}=1.5$.
Full binary calculations \cite{han02} demonstrate that
$q_{\rm crit} \sim 1.2$ is typical for RLOF in FGB stars.
We do not explicitly include tidally
enhanced stellar winds \cite{tou88,egg89a,han95c} in our calculation.
Using a larger value for $q_{\rm crit}$ is
equivalent to including a tidally enhanced stellar wind
to some degree while keeping the simulation simple
(see HPMM for details). Alternatively, we also
adopt $q_{\rm crit}=1.2$ in order to examine the consequences of
varying this important criterion.

For stable RLOF, we assume that a fraction $\alpha_{\rm RLOF}$ of the
mass lost from the primary is transferred onto the gainer, while the
rest is lost from the system ($\alpha_{\rm RLOF}=1$ means that RLOF is
conservative).  Note, however, that we assume that
mass transfer
is always conservative when the donor is a MS star.  
The mass lost from the system also takes away a specific angular 
momentum
$\alpha$ in units of the specific angular momentum of the system. The
unit is expressed as $2\pi a^2/P$, where $a$ is the separation and $P$
is the orbital period of the binary (see Podsiadlowski, Joss \& Hsu
1992 for details).  Stable RLOF usually results in a wide WD
binary. Some of the wide WD binaries may contain hot 
subdwarf stars
and MS companions if RLOF occurs near the tip of the FGB ({\it 1st
stable RLOF channel} for the formation of hot subdwarfs).  In order to
reproduce Galactic hot subdwarfs, we use $\alpha_{\rm RLOF}=0.5$ and
$\alpha=1$ for the first stable RLOF in all systems except for those
on the MS (see HPMM for details).

If RLOF is dynamically unstable, a common envelope (CE) 
may be formed \cite{pac76}, and
if the orbital energy deposited in the envelope can overcome its
binding energy, the CE may be ejected. For the CE ejection criterion,
we introduced two model parameters, $\alpha_{\rm CE}$ for the common
envelope ejection efficiency and $\alpha_{\rm th}$ for the thermal
contribution to the binding energy of the envelope, which we write as
\begin{equation}
\alpha_{\rm CE}\, \Delta E_{\rm orb} > E_{\rm gr} - \alpha_{\rm th}
\,E_{\rm th},
\end{equation}
where $\Delta E_{\rm orb}$ is the orbital energy that is released,
$E_{\rm gr}$ is the gravitational binding energy and $E_{\rm th}$ is
the thermal energy of the envelope.  Both $E_{\rm gr}$ and $E_{\rm
th}$ are obtained from full stellar structure calculations (see Han,
Podsiadlowski \& Eggleton 1994 for details; also see
Dewi \& Tauris 2000) instead of analytical approximations.  CE
ejection leads to the formation of a close WD binary.
Some of the close WD binaries may contain hot subdwarf stars and MS
companions if the CE occurs near the tip of the FGB
({\it 1st CE channel} for the formation of hot subdwarfs).
We adopt $\alpha_{\rm CE}=\alpha_{\rm th}=0.75$ in our standard model,
and $\alpha_{\rm CE}=\alpha_{\rm th}=1.0$ to investigate the
effect of varying the CE ejection parameters.

The WD binary formed from the first RLOF phase continues to evolve,
and the secondary may fill its Roche lobe as a red giant. The system
then experiences a second RLOF phase.  If the mass ratio at the onset
of RLOF is greater than $q_{\rm crit}$ given in
table~3 of Han \etal \shortcite{han02},
RLOF is dynamically unstable, leading again to a
CE phase.  If the CE is ejected, a hot subdwarf star may be formed.
The hot sub\-dwarf binary has a short orbital period and a WD companion
({\it 2nd CE channel} for the formation of hot subdwarfs).
However, RLOF may be stable if the mass ratio is sufficiently small.
In this case, we assume that the mass lost from the mass donor is all
lost from the system, carrying away the same specific angular momentum
as pertains to the WD companion. Stable RLOF may then result in the
formation of a hot subdwarf binary with a WD companion and a long orbital
period (typically $\sim 1000\,{\rm d}$, {\it 2nd stable RLOF channel}
for the formation of hot subdwarfs).

If the second RLOF phase results in a CE phase and the CE is ejected,
a double white dwarf system is formed \cite{web84,ibe86,han98}.  Some
of the double WD systems contain two helium WDs.  Angular momentum
loss due to gravitational wave radiation may then cause the shrinking of
the orbital separation until the less massive white dwarf starts to
fill its Roche lobe. This will lead to its dynamical disruption if
\begin{equation}
q \ga 0.7 - 0.1 (M_2/\,M_\odot)
\end{equation}
or $M_1\ga 0.3\,M_\odot$, where $M_1$ is the mass of the donor
(i.e. the less massive WD) and $M_2$ is the mass of the gainer
\cite{han99}. This is expected to always lead to a complete merger of
the two white dwarfs. The merger can also produce a hot
subdwarf star, but in
this case the hot subdwarf star is a single object
({\it He WD merger channel} for the formation of hot subdwarfs).
If the lighter WD is not
disrupted, RLOF is stable and an AM CVn system is formed.

In this paper, we do not include a tidally enhanced stellar wind
explicitly as was done in Han \etal (1995) and Han (1998).  Instead
we use a standard Reimers wind formula (Reimers 1975) with $\eta =
1/4$ \cite{ren81,ibe83,car96} which is included in our stellar
models. This is to keep the simulations as simple as possible,
although the effects of a tidally enhanced wind can to some degree be
implicitly included by using a larger value of $q_{\rm crit}$.  We
also employ a standard magnetic braking law \cite{ver81,rap83} where
appropriate (see Podsiadlowski, Han \& Rappaport 2002 for details and
further discussion).

\subsection{Monte-Carlo simulation parameters}

In order to investigate the UV-upturn phenomenon due to hot subdwarfs,
we have performed a series of Monte-Carlo simulations where
we follow the evolution of a sample of a million binaries
(single stars are in effect treated as very wide binaries that
do not interact with each other),
including the hot subdwarfs produced in the simulations,
according to our grids of stellar models.
The simulations require as input the star formation rate (SFR),
the initial mass function (IMF) of the primary,
the initial mass-ratio distribution and
the distribution of initial orbital separations.

(1) The SFR is taken to be a single starburst in most of our
simulations; all the stars
have the same age ($t_{\rm SSP}$) and the same metallicity ($Z=0.02$), and
constitute a simple stellar population (SSP). In some simulations
a composite stellar population (CSP) is also used (Section~3.3).

(2) A simple approximation to the IMF of Miller \& Scalo \shortcite{mil79}
is used; the primary mass is generated with the formula of Eggleton, 
Fitchett
\& Tout \shortcite{egg89b},
\begin{equation}
M_1={0.19X\over (1-X)^{0.75}+0.032(1-X)^{0.25}},
\end{equation}
where $X$ is a random number uniformly distributed between 0 and 1.
The adopted range of primary masses is 0.8 to $100.0\,M_\odot$. The
studies by Kroupa, Tout
\& Gilmore \shortcite{kro93} and Zoccali \etal \shortcite{zoc00}
support this IMF.

(3) The mass-ratio distribution is quite uncertain. We mainly
take a constant mass-ratio distribution
\cite{maz92,gol94,hea95,hal98,sha02},
\begin{equation}
n(q')=1,\qquad  0\leq q' \leq 1,
\end{equation}
where $q'=1/q=M_2/M_1$. As alternatives we also consider a rising mass ratio
distribution
\begin{equation}
n(q')=2q',\qquad  0\leq q' \leq 1,
\end{equation}
and the case where both binary components are chosen randomly and
independently from the same IMF.

(4) We assume that all stars are members of binary systems and that the
distribution of separations is constant in $\log a$
(where $a$ is the orbital separation) for wide
binaries and falls off smoothly at close separations:
\begin{equation}
an(a)=\cases {\alpha_{\rm sep}({a\over a_0})^m, & $a\leq a_0$;\cr
               \alpha_{\rm sep}, & $a_0 < a < a_1$,\cr }
\end{equation}
where $\alpha_{\rm sep} \approx 0.070$, $a_0=10\,R_\odot$,
$a_1=5.75\times 10^6\,R_\odot=0.13\,{\rm pc}$, and $m\approx 1.2$.
This distribution implies that there is an equal number of wide binary
systems per logarithmic interval and that approximately 50 per cent
of stellar systems are binary systems with orbital periods less than
100\,yr.

\subsection{Spectral library}

In order to obtain the colours and the spectral energy distribution
(SED) of the populations produced by our simulations, we have
calculated spectra for hot subdwarfs using plane-parallel static model
stellar atmospheres computed with the {\scriptsize ATLAS9} stellar
atmosphere code \cite{kur92} with the assumption of local
thermodynamic equilibrium (LTE).  Solar metal abundances were adopted
\cite{and89} and line blanketing is approximated by appropriate opacity
distribution functions interpolated for the chosen helium
abundance. The resulting model atmosphere grid covers a wide range of
effective temperatures ($10,000 \le T_{\rm eff} \le 40, 000{\rm K}$
with a spacing of $\Delta T=1000{\rm K}$), gravities ($5.0\le \log g
\le 7.0$ with $\Delta \log g=0.2$), and helium abundances ($-3 \le
[He/H] \le 0$), as appropriate for the hot subdwarfs produced in the
BPS code. For the spectrum and colours of other single stars, we use
the latest version of the comprehensive BaSeL library of theoretical
stellar spectra (see Lejeune \etal 1997, 1998 for a description),
which gives the colours and SEDs of stars with a wide range of
$Z$, $\log g$ and $T_{\rm eff}$.

\subsection{Observables from the model}

Our model follows the evolution of the integrated spectra
of stellar populations. In order to compare the results with observations,
we calculate the following observables as well as $UBV$ colours
in the Johnson system \cite{joh53}.

\begin{enumerate}

\item $(1550-V)$ is a colour defined by BBBFL. It is a combination
of the short-wavelength IUE flux and the $V$ magnitude
and is used to express the magnitude of the UV-upturn:
\begin{equation}
(1550-V)=-2.5\log (f_{\lambda, 1250-1850}/f_{\lambda, 5055-5945}),
\end{equation}
where $f_{\lambda, 1250-1850}$ is the energy
flux per unit wavelength averaged between 1250 and 1850\AA~
and $f_{\lambda, 5055-5945}$ the flux averaged between 5055 and
5945\AA~(for the $V$ band):

\item $(1550-2500)$ is a colour defined for the IUE flux
by Dorman, O'Connell \& Rood
\shortcite{dor95}.
\begin{equation}
(1550-2500)=-2.5\log (f_{\lambda, 1250-1850}/f_{\lambda, 2200-2800}).
\end{equation}

\item $(2000-V)$ is a colour used by Deharveng, Boselli \& Donas
\shortcite{deh02} in their study of UV properties of the early-type galaxies
observed with the balloon-borne FOCA experiment \cite{don90,don95}:
\begin{equation}
(2000-V)=-2.5\log (f_{\lambda, 1921-2109}/f_{\lambda, 5055-5945}).
\end{equation}

\item $(FUV-NUV)_{\rm AB}$, $(FUV-r)_{\rm AB}$, $(NUV-r)_{\rm AB}$
are colours from GALEX and SDSS.
GALEX, a NASA Small Explorer mission, has two bands in
its ultraviolet (UV) survey: a far-UV band centered on 1530\,\AA~ and
a near-UV band centered on  2310\,\AA~ \cite{mar05,ric05},
while SDSS has five passbands,
u at 3551\,\AA, g at 4686\,\AA, r at 6165\,\AA,
i at 7481\,\AA, and z at 8931\,\AA~ \cite{fuk96,gun98,smi02}.
The magnitudes are in
the AB system of Oke \& Gunn \shortcite{oke83}:
\begin{equation}
(FUV-NUV)_{\rm AB}=-2.5\log (f_{\nu, 1350-1750}/f_{\nu, 1750-2750}),
\end{equation}
\begin{equation}
(FUV-r)_{\rm AB}=-2.5\log (f_{\nu, 1350-1750}/f_{\nu, 5500-7000}),
\end{equation}
\begin{equation}
(NUV-r)_{\rm AB}=-2.5\log (f_{\nu, 1750-2750}/f_{\nu, 5500-7000}),
\end{equation}
where $f_{\nu, 1350-1750}$, $f_{\nu, 1750-2750}$, $f_{\nu, 5500-7000}$
are the energy fluxes per unit frequency averaged in the
frequency bands corresponding to
1350 and 1750\,\AA, 1750 and 2750\,\AA, 5500 and 7000\,\AA,
respectively.

\item
$\beta_{\rm FUV}$ is a far-UV spectral index we defined to measure
the SED slope between 1075 and 1750\,\AA:
\begin{equation}
f_\lambda \sim \lambda^{\beta_{\rm FUV}}, \qquad  1075 < \lambda < 1750
{\,\rm\AA,}
\label{slope}
\end{equation}
where $f_\lambda$ is the energy flux per unit
wavelength. In this paper, we fit far-UV SEDs with
equation (\ref{slope}) to obtain $\beta_{\rm FUV}$.
In the fitting we ignored the part between 1175 and 1250\AA~
for the theoretical SEDs from our model,
as this part corresponds to the L$_\alpha$ line.
We also obtained $\beta_{\rm FUV}$ via a similar fitting
for early-type galaxies observed with the HUT
\cite{bro97,bro04} and the IUE \cite{bur88}.
We again ignored the part between 1175 and 1250\AA~ for the HUT SEDs,
while the IUE SEDs do not have any data points below 1250\AA.

\end{enumerate}

\section{Simulations}

\begin{table}
\caption{Simulation sets (metallicity $Z=0.02$)}
\begin{tabular}{llllll}
\hline
&Set & $n(q')$ & $q_{\rm c}$ & $\alpha_{\rm CE}$ & $\alpha_{\rm th}$ \\
\hline
&& &&&\\
\multicolumn{6}{l}{Standard SSP simulation set with $t_{\rm SSP}$ varying
upto 15\,Gyr}\\
&1   & constant & 1.5 & 0.75 & 0.75\\
&& &&&\\
\multicolumn{6}{l}{SSP simulation sets with varying model parameters}\\
&2   & uncorrelated & 1.5 & 0.75 & 0.75\\
&3   & rising & 1.5 & 0.75 & 0.75\\
&4   & constant & 1.2 & 0.75 & 0.75\\
&5   & constant & 1.5 & 1.0 & 1.0\\
&& &&&\\
\multicolumn{6}{l}{CSP simulation sets with a $t_{\rm major}$ and
variable $t_{\rm minor}$ and $f$}\\
&6   & constant & 1.5 & 0.75 & 0.75\\
\hline
\end{tabular}

\medskip
   Note -  $n(q')$ = initial mass-ratio distribution;
   $q_{\rm c}$ = the critical mass ratio above which the
    first RLOF on the FGB or AGB is unstable;
   $\alpha_{\rm CE}$ = CE ejection efficiency;
   $\alpha_{\rm th}$ = thermal contribution to CE ejection;
   $t_{\rm SSP}$ = the age of a SSP;
   $t_{\rm major}$ = the age of the major population in a CSP;
   $t_{\rm minor}$ = the age of the minor population in a CSP;
   $f$ = the ratio of the mass of the minor population
   to the total mass in a CSP.
\label{models}
\end{table}

In order to investigate the UV-upturn systematically,
we performed six sets of simulations for a Population I composition
($X=0.70$, $Y=0.28$ and $Z=0.02$).
The first set is a standard set with the best-choice model parameters
from HPMM. In Sets 2 to 5 we systematically vary
the model parameters, and Set 6 models a
composite stellar population
(Table~\ref{models}).

\subsection{Standard simulation set}

In the HPMM model, hot subdwarfs are produced through binary interactions by
stable RLOF, CE ejection or He WD mergers.  
The main parameters in the HPMM model are: $n(q')$, the initial mass
ratio distribution, $q_{\rm c}$, the critical mass ratio above which
the 1st RLOF on the FGB or AGB is unstable, $\alpha_{\rm CE}$, the CE
ejection efficiency parameter, and $\alpha_{\rm th}$, the contribution of
thermal energy to the binding energy of the CE (see Sections~2.1 and
2.2 for details).  The model that best reproduces the observed properties
of hot subdwarfs in our Galaxy has
$n(q')=1$, $q_{\rm c}=1.5$, $\alpha_{\rm CE}=0.75$, $\alpha_{\rm
th}=0.75$, (see section 7.4 of Han \etal 2003). These are the parameters
adopted in our standard simulation.

In our standard set, we first construct a SSP containing a million
binaries (Section~2.2). The binaries are formed simultaneously
(i.e. in a single starburst) and with the same metallicity
($Z=0.02$). The SSP is evolved with the BPS code (Section~2.1),
and the results are convolved with our spectral libraries
(Section~2.3) to produce integrated SEDs and other observables.  The
SEDs are normalised to a stellar population of mass $10^{10}M_\odot$
at a distance of 10\,Mpc.

\subsection{Simulation sets with varying model parameters}

In order to investigate the importance of the model parameters, we
also carried out simulation sets in which we systematically varied the
main parameters.  Specifically, we adopted three initial mass-ratio
distributions: a constant mass-ratio distribution, a rising mass-ratio
distribution, and one where the component masses are uncorrelated and
drawn independently from a Miller-Scalo IMF (Section~2.2).  We also
varied the value of $q_{\rm c}$ in the instability criterion for the
first RLOF phase on the FGB or AGB from 1.5 to 1.2, the parameter
$\alpha_{\rm CE}$ for CE ejection efficiency and the parameter
$\alpha_{\rm th}$ for the thermal contribution to CE ejection from
0.75 to 1.0.

\subsection{Simulation sets for composite stellar populations}

Many early-type galaxies show evidence for some moderate amount of
recent star formation \cite{yi05,sal05,sch06}.  Therefore, we also
perform simulations in which we evolve composite stellar
populations. Here, a composite stellar population (CSP) consists of
two populations, a major old one and a minor younger one.  The major
population has solar metallicity and an age of $t_{\rm major}$, where
all stars formed in a single burst $t_{\rm major}$ ago, while the
minor one has solar metallicity and an age $t_{\rm minor}$, where all
stars formed in a starburst starting $t_{\rm minor}$ ago and lasting
0.1\,Gyr.  The minor population fraction $f$ is the ratio of the mass
of the minor population to the total mass of the CSP (for $f=100\%$
the CSP is actually a SSP with an age $t_{\rm minor}$).

\section{Results and Discussion}

\subsection{Simple stellar populations}

\subsubsection{Evolution of the integrated SED}

\begin{figure}
\epsfig{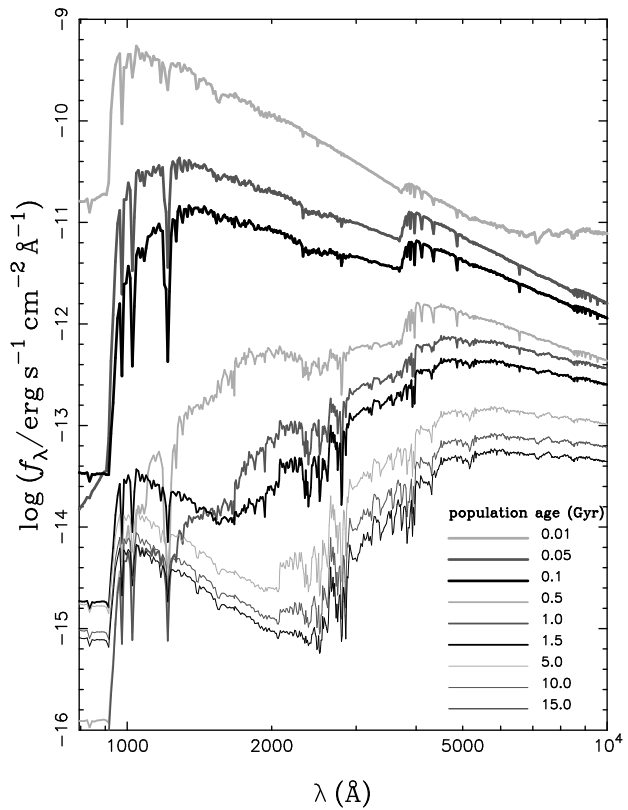}
\caption{ The evolution of the restframe intrinsic spectral energy
distribution (SED) for a simulated galaxy in which all stars formed at
the same time, representing a simple stellar population (SSP).  The
stellar population (including binaries) has a mass of $10^{10}M_\odot$
and the galaxy is assumed to be at a distance of 10\,Mpc.  The figure
is for the standard simulation set, and no offset has been applied to the
SEDs.  }
\label{sed}
\end{figure}

In our standard simulation set, we follow the
evolution of the integrated SED of a SSP (including binaries)
of $10^{10}M_\odot$ up to $t_{\rm SSP}=15\,{\rm Gyr}$.
The SSP is assumed to be at a distance of 10\,Mpc and
the evolution is shown in Figure~\ref{sed}.
Note that hot subdwarfs originating from binary interactions
start to dominate the far-UV after $\sim 1$\,Gyr.

We have compiled a file containing the spectra of the SSP with ages from
0.1\,Gyr to 15\,Gyr and devised a small FORTRAN code to read the file (and
to plot the SEDs with PGPLOT). The file and the code are available
online\footnote{The file and the code are available
on the VizieR data base of the astronomical catalogues at the Centre
de Donn\'ees astronomiques de Strasbourg (CDS) web site
(http://cdsarc.u-strasbg.fr/) and on ZH's personal website
(http://www.zhanwenhan.com/download/uv-upturn.html).}.
In order to be able to apply the model directly,
we have also provided in the file
the spectra of the SSP without any binary interactions
considered. This provides an easy way to examine the differences
in the spectra for simulations with and without binary interactions.

\subsubsection{Evolution of the UV-upturn}
\begin{table*}
\caption{Colour evolution of a simple stellar population (including
          binaries) of $10^{10}M_\odot$ for the standard simulation set.
      This table is also available in machine-readable form
     on the VizieR data base of the astronomical catalogues at the Centre
     de Donn\'ees astronomiques de Strasbourg (CDS) web site
     (http://cdsarc.u-strasbg.fr/) and on ZH's personal website
     (http://www.zhanwenhan.com/download/uv-upturn.html).
}
\begin{tabular}{rrrrrrrrrrr}
\hline
  $\log (t_{\rm SSP})$ & $M_{\rm V}$ & $B-V$ & $15-V$
  & $20-V$ & $25-V$ & $15-25$
  & $FUV-r$ & $NUV-r$ & $FUV-NUV$
  & $\beta_{\rm FUV}$\\
\hline
  -1.000&-22.416&  0.173& -1.274& -0.953& -0.447& -0.827&  1.483&  1.296&  
0.188&  0.775\\
  -0.975&-22.346&  0.165& -1.245& -0.938& -0.432& -0.813&  1.503&  1.304&  
0.199&  0.941\\
  -0.950&-22.306&  0.175& -1.195& -0.901& -0.396& -0.799&  1.555&  1.346&  
0.209&  1.100\\
  -0.925&-22.289&  0.185& -1.106& -0.829& -0.326& -0.780&  1.644&  1.421&  
0.223&  1.305\\
  -0.900&-22.232&  0.194& -1.053& -0.788& -0.286& -0.767&  1.699&  1.466&  
0.233&  1.479\\
  -0.875&-22.168&  0.192& -1.019& -0.766& -0.265& -0.754&  1.730&  1.487&  
0.243&  1.659\\
  -0.850&-22.135&  0.199& -0.959& -0.717& -0.215& -0.744&  1.792&  1.542&  
0.250&  1.842\\
  -0.825&-22.053&  0.199& -0.969& -0.723& -0.216& -0.753&  1.780&  1.537&  
0.243&  1.968\\
  -0.800&-22.000&  0.200& -0.945& -0.701& -0.189& -0.756&  1.802&  1.561&  
0.241&  2.182\\
  -0.775&-21.949&  0.205& -0.892& -0.660& -0.148& -0.744&  1.855&  1.605&  
0.250&  2.407\\
\\
  -0.750&-21.923&  0.207& -0.789& -0.587& -0.086& -0.703&  1.955&  1.674&  
0.281&  2.656\\
  -0.725&-21.885&  0.211& -0.682& -0.514& -0.025& -0.658&  2.058&  1.743&  
0.316&  2.935\\
  -0.700&-21.839&  0.218& -0.569& -0.437&  0.041& -0.609&  2.169&  1.816&  
0.353&  3.314\\
  -0.675&-21.804&  0.221& -0.468& -0.367&  0.101& -0.569&  2.266&  1.882&  
0.384&  3.565\\
  -0.650&-21.767&  0.223& -0.377& -0.307&  0.152& -0.529&  2.355&  1.939&  
0.416&  3.832\\
  -0.625&-21.732&  0.230& -0.272& -0.235&  0.215& -0.486&  2.459&  2.009&  
0.451&  4.087\\
  -0.600&-21.689&  0.235& -0.166& -0.167&  0.273& -0.439&  2.565&  2.075&  
0.490&  4.343\\
  -0.575&-21.644&  0.242& -0.051& -0.092&  0.337& -0.388&  2.681&  2.148&  
0.532&  4.583\\
  -0.550&-21.606&  0.246&  0.074& -0.015&  0.403& -0.329&  2.806&  2.222&  
0.583&  4.879\\
  -0.525&-21.546&  0.250&  0.188&  0.048&  0.456& -0.268&  2.918&  2.280&  
0.638&  5.176\\
\\
  -0.500&-21.512&  0.263&  0.315&  0.122&  0.522& -0.207&  3.055&  2.359&  
0.695&  5.384\\
  -0.475&-21.478&  0.274&  0.444&  0.196&  0.587& -0.144&  3.189&  2.433&  
0.757&  5.503\\
  -0.450&-21.424&  0.285&  0.556&  0.254&  0.637& -0.081&  3.316&  2.495&  
0.820&  5.524\\
  -0.425&-21.377&  0.292&  0.688&  0.322&  0.694& -0.005&  3.458&  2.560&  
0.898&  5.603\\
  -0.400&-21.335&  0.308&  0.872&  0.421&  0.777&  0.095&  3.661&  2.660&  
1.001&  5.881\\
  -0.375&-21.290&  0.319&  1.015&  0.497&  0.840&  0.175&  3.825&  2.736&  
1.089&  5.642\\
  -0.350&-21.253&  0.340&  1.172&  0.591&  0.919&  0.253&  4.011&  2.834&  
1.176&  5.545\\
  -0.325&-21.217&  0.362&  1.374&  0.706&  1.015&  0.360&  4.246&  2.951&  
1.295&  5.507\\
  -0.300&-21.162&  0.375&  1.543&  0.795&  1.084&  0.459&  4.440&  3.036&  
1.404&  5.764\\
  -0.275&-21.121&  0.394&  1.723&  0.903&  1.168&  0.555&  4.651&  3.143&  
1.507&  5.827\\
\\
  -0.250&-21.068&  0.410&  1.897&  1.012&  1.245&  0.653&  4.846&  3.242&  
1.604&  6.004\\
  -0.225&-21.028&  0.434&  2.116&  1.152&  1.344&  0.772&  5.095&  3.374&  
1.721&  5.839\\
  -0.200&-20.994&  0.462&  2.339&  1.305&  1.448&  0.890&  5.343&  3.517&  
1.826&  5.968\\
  -0.175&-20.965&  0.495&  2.597&  1.485&  1.571&  1.026&  5.629&  3.682&  
1.947&  5.989\\
  -0.150&-20.937&  0.524&  2.778&  1.632&  1.671&  1.107&  5.820&  3.816&  
2.004&  6.248\\
  -0.125&-20.930&  0.572&  2.964&  1.814&  1.804&  1.159&  6.018&  3.994&  
2.024&  6.113\\
  -0.100&-20.864&  0.590&  3.123&  1.932&  1.873&  1.250&  6.184&  4.092&  
2.091&  5.806\\
  -0.075&-20.801&  0.603&  3.274&  2.038&  1.933&  1.340&  6.336&  4.174&  
2.162&  5.673\\
  -0.050&-20.730&  0.622&  3.440&  2.142&  1.992&  1.448&  6.522&  4.264&  
2.258&  5.434\\
  -0.025&-20.664&  0.639&  3.655&  2.254&  2.056&  1.599&  6.768&  4.355&  
2.413&  4.681\\
\\
   0.000&-20.583&  0.649&  3.771&  2.336&  2.096&  1.675&  6.897&  4.415&  
2.482&  3.445\\
   0.025&-20.511&  0.666&  3.822&  2.424&  2.144&  1.677&  6.955&  4.488&  
2.467&  2.006\\
   0.050&-20.419&  0.672&  3.737&  2.484&  2.171&  1.566&  6.860&  4.528&  
2.332&  0.404\\
   0.075&-20.313&  0.666&  3.569&  2.517&  2.176&  1.393&  6.672&  4.534&  
2.137& -0.684\\
   0.100&-20.259&  0.685&  3.420&  2.607&  2.241&  1.179&  6.509&  4.612&  
1.897& -1.360\\
   0.125&-20.201&  0.707&  3.441&  2.738&  2.323&  1.118&  6.536&  4.719&  
1.817& -1.734\\
   0.150&-20.150&  0.722&  3.547&  2.866&  2.398&  1.149&  6.646&  4.814&  
1.832& -1.868\\
   0.175&-20.081&  0.731&  3.603&  2.974&  2.455&  1.148&  6.704&  4.888&  
1.816& -2.019\\
   0.200&-20.028&  0.750&  3.642&  3.094&  2.525&  1.117&  6.748&  4.983&  
1.766& -2.148\\
   0.225&-19.973&  0.766&  3.731&  3.240&  2.606&  1.125&  6.837&  5.085&  
1.752& -2.270\\
\hline
\end{tabular}
\label{colours}
\end{table*}

\setcounter{table}{1}
\begin{table*}
\caption{continued}
\begin{tabular}{rrrrrrrrrrr}
\hline
  $\log (t_{\rm SSP})$ & $M_{\rm V}$ & $B-V$ & $15-V$
  & $20-V$ & $25-V$ & $15-25$
  & $FUV-r$ & $NUV-r$ & $FUV-NUV$
  & $\beta_{\rm FUV}$\\
\hline
   0.250&-19.906&  0.773&  3.804&  3.359&  2.664&  1.140&  6.902&  5.157&  
1.745& -2.342\\
   0.275&-19.853&  0.790&  3.886&  3.492&  2.735&  1.151&  6.992&  5.254&  
1.739& -2.457\\
   0.300&-19.789&  0.799&  3.936&  3.571&  2.781&  1.155&  7.042&  5.311&  
1.731& -2.466\\
   0.325&-19.746&  0.811&  4.005&  3.684&  2.842&  1.162&  7.115&  5.389&  
1.726& -2.573\\
   0.350&-19.664&  0.812&  4.010&  3.761&  2.870&  1.140&  7.119&  5.428&  
1.691& -2.680\\
   0.375&-19.588&  0.815&  4.007&  3.821&  2.895&  1.111&  7.113&  5.462&  
1.651& -2.706\\
   0.400&-19.533&  0.818&  3.968&  3.879&  2.932&  1.036&  7.067&  5.499&  
1.568& -2.759\\
   0.425&-19.484&  0.827&  3.922&  3.915&  2.968&  0.954&  7.025&  5.539&  
1.486& -2.787\\
   0.450&-19.412&  0.830&  3.854&  3.931&  2.987&  0.867&  6.957&  5.557&  
1.401& -2.828\\
   0.475&-19.374&  0.846&  3.856&  4.009&  3.049&  0.807&  6.965&  5.629&  
1.336& -2.865\\
\\
   0.500&-19.337&  0.857&  3.817&  4.041&  3.096&  0.720&  6.924&  5.673&  
1.251& -2.851\\
   0.525&-19.288&  0.869&  3.743&  4.041&  3.127&  0.615&  6.857&  5.703&  
1.154& -2.857\\
   0.550&-19.201&  0.864&  3.653&  4.026&  3.128&  0.525&  6.763&  5.693&  
1.070& -2.896\\
   0.575&-19.170&  0.879&  3.666&  4.082&  3.189&  0.477&  6.782&  5.758&  
1.024& -2.914\\
   0.600&-19.089&  0.876&  3.654&  4.103&  3.203&  0.451&  6.767&  5.770&  
0.996& -2.930\\
   0.625&-19.055&  0.891&  3.708&  4.188&  3.271&  0.437&  6.828&  5.850&  
0.978& -2.944\\
   0.650&-18.992&  0.896&  3.666&  4.179&  3.290&  0.376&  6.785&  5.862&  
0.923& -2.925\\
   0.675&-18.957&  0.904&  3.695&  4.230&  3.340&  0.355&  6.817&  5.914&  
0.903& -2.933\\
   0.700&-18.880&  0.905&  3.667&  4.239&  3.357&  0.310&  6.788&  5.928&  
0.860& -2.955\\
   0.725&-18.812&  0.906&  3.635&  4.238&  3.371&  0.264&  6.754&  5.936&  
0.818& -2.962\\
\\
   0.750&-18.758&  0.913&  3.631&  4.247&  3.402&  0.229&  6.754&  5.966&  
0.788& -2.944\\
   0.775&-18.710&  0.925&  3.682&  4.310&  3.453&  0.229&  6.814&  6.029&  
0.785& -2.958\\
   0.800&-18.680&  0.936&  3.676&  4.325&  3.502&  0.174&  6.810&  6.071&  
0.739& -2.956\\
   0.825&-18.576&  0.929&  3.615&  4.290&  3.492&  0.123&  6.745&  6.050&  
0.695& -2.964\\
   0.850&-18.556&  0.947&  3.606&  4.314&  3.560&  0.046&  6.744&  6.113&  
0.631& -2.974\\
   0.875&-18.473&  0.945&  3.629&  4.348&  3.583&  0.046&  6.765&  6.138&  
0.627& -2.979\\
   0.900&-18.429&  0.959&  3.591&  4.338&  3.626& -0.035&  6.735&  6.173&  
0.562& -2.980\\
   0.925&-18.362&  0.958&  3.552&  4.319&  3.647& -0.095&  6.691&  6.176&  
0.514& -2.988\\
   0.950&-18.339&  0.981&  3.613&  4.391&  3.729& -0.116&  6.767&  6.271&  
0.496& -2.986\\
   0.975&-18.236&  0.972&  3.528&  4.322&  3.711& -0.184&  6.675&  6.231&  
0.445& -2.992\\
\\
   1.000&-18.191&  0.982&  3.570&  4.374&  3.768& -0.198&  6.724&  6.292&  
0.432& -2.989\\
   1.025&-18.100&  0.975&  3.499&  4.320&  3.764& -0.265&  6.645&  6.263&  
0.382& -2.998\\
   1.050&-18.100&  0.995&  3.481&  4.321&  3.843& -0.362&  6.634&  6.321&  
0.313& -3.005\\
   1.075&-18.036&  1.004&  3.514&  4.359&  3.898& -0.384&  6.672&  6.376&  
0.296& -3.002\\
   1.100&-18.044&  1.032&  3.541&  4.396&  3.997& -0.456&  6.712&  6.465&  
0.247& -2.998\\
   1.125&-17.947&  1.029&  3.473&  4.338&  3.998& -0.525&  6.641&  6.441&  
0.201& -2.999\\
   1.150&-17.884&  1.033&  3.473&  4.345&  4.039& -0.566&  6.641&  6.468&  
0.173& -2.995\\
   1.175&-17.800&  1.030&  3.463&  4.338&  4.062& -0.600&  6.626&  6.475&  
0.151& -2.995\\
\hline
\end{tabular}

\medskip
   Note -
   $t_{\rm SSP}$ = population age in Gyr;
   $M_{\rm V}$ = absolute visual magnitude;
   $B-V$ = $(B-V)$;
   $15-V$ = $(1550-V)$;
   $20-V$ = $(2000-V)$;
   $25-V$ = $(2500-V)$;
   $15-25$ = $(1550-2500)$;
   $FUV-r$ = $(FUV-r)_{\rm AB}$;
   $NUV-r$ = $(FUV-r)_{\rm AB}$;
   $FUV-NUV$ = $(FUV-NUV)_{\rm AB}$;
   $\beta_{\rm FUV}$ = far-UV spectral index.\hfill
\end{table*}

\begin{figure}
\epsfig{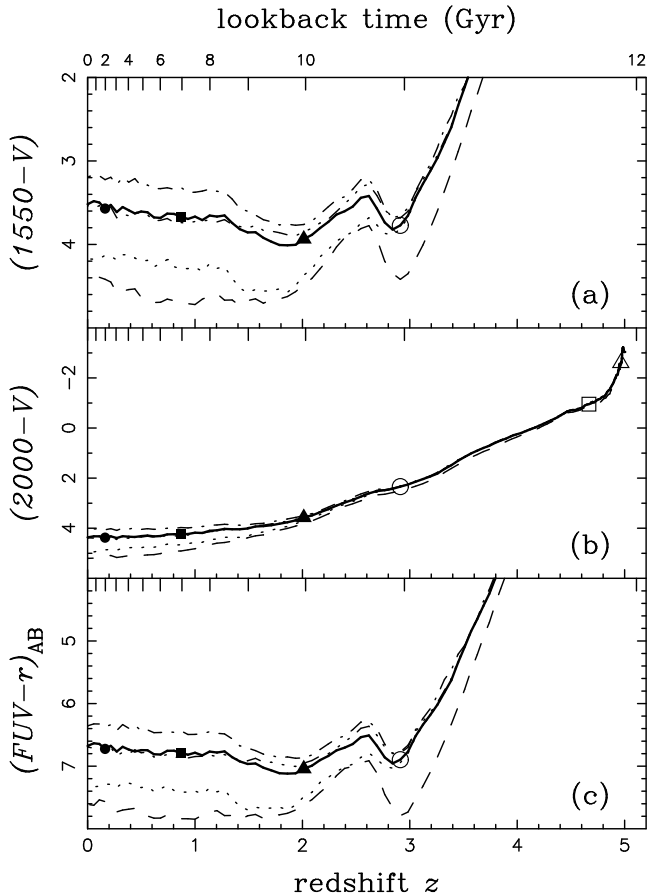}
\caption{
The evolution of restframe intrinsic colours
$(1550-V)$, $(2000-V)$ and $(FUV-r)_{\rm AB}$
with redshift (lookback time) for a simple stellar
population (including binaries). Solid, dashed, dash-dotted,
dotted, dash-dot-dot-dot curves are for simulation sets
1 (standard set), 2, 3, 4, 5, respectively.
Ages are denoted by open triangles (0.01\,Gyr),
open squares (0.1\,Gyr), open circles (1\,Gyr),
filled triangles (2\,Gyr), filled squares (5\,Gyr) and
filled circles (10\,Gyr).
}
\label{uvz1}
\end{figure}

\begin{figure}
\epsfig{file=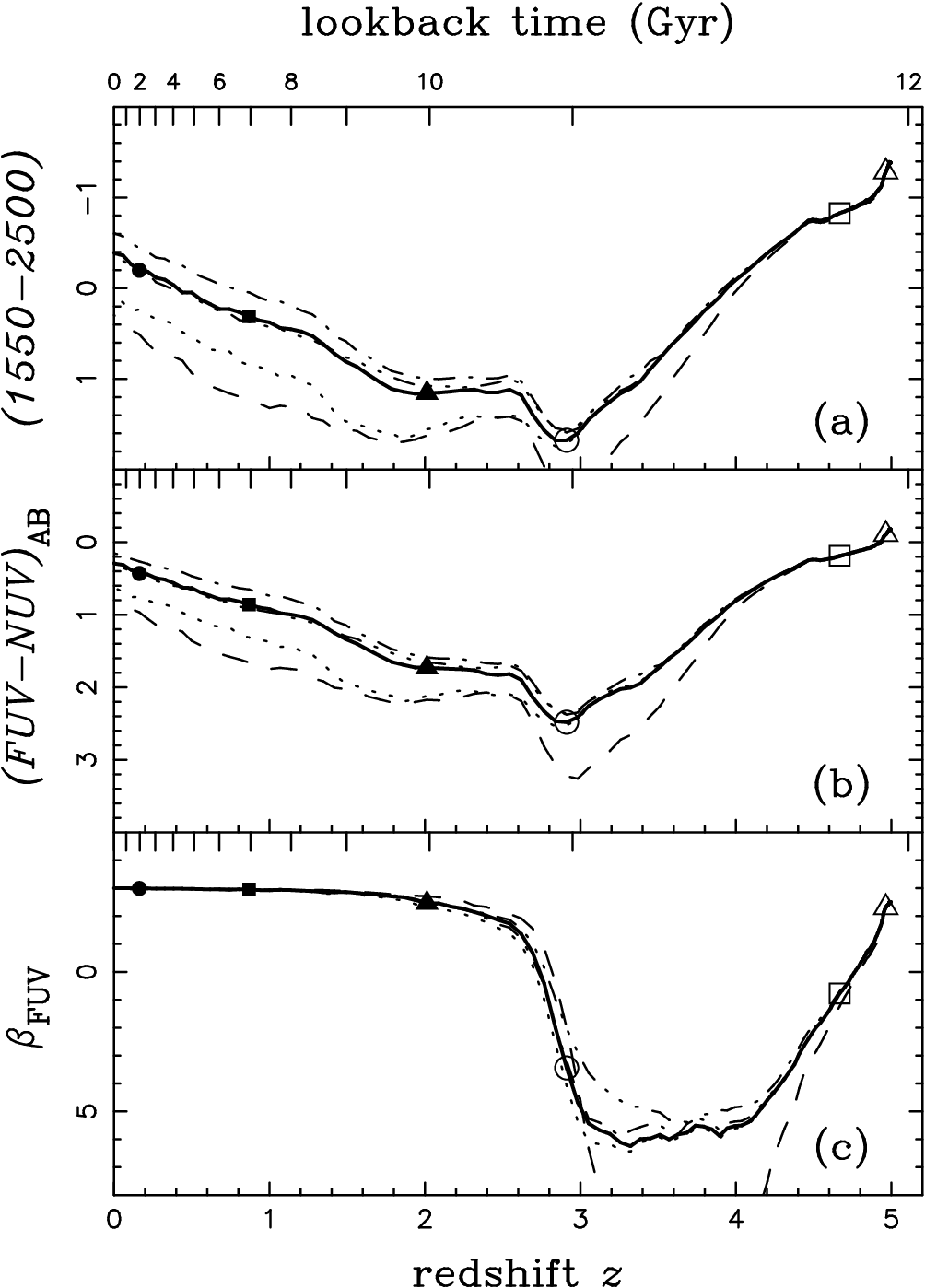,width=8.5cm}
\caption{
Similar to Figure~\ref{uvz1}, but for colours $(1550-2500)$,
$(FUV-NUV)_{\rm AB}$, and far-UV spectral index $\beta_{\rm FUV}$.
}
\label{uvz2}
\end{figure}

The colours of the SSP evolve in time, and Table~\ref{colours} lists
the colours of a SSP (including binaries) of $10^{10}M_\odot$
at various ages for the standard simulation set.
In order to see how the colours evolve with redshift,
we adopted a $\Lambda$CDM cosmology \cite{car92} with
cosmological parameters of
$H_0=72{\rm km/s/Mpc}$, $\Omega_{\rm M}=0.3$ and $\Omega_\Lambda=0.7$,
and assumed a star-formation redshift of $z_{\rm f}=5$ to
obtain Figures~\ref{uvz1} and ~\ref{uvz2}.
The figures show the evolution of the restframe intrinsic colours
and the evolution of the far-UV spectral index with redshift (lookback 
time).
As these figures show, the UV-upturn does not evolve much with redshift;
for an old stellar population (i.e. with a redshift $z\sim 0$
or an age of $\sim 12$\,Gyr),
$(1550-V)\sim 3.5$, $(UV-V)\sim 4.4$,
$(FUV-r)_{\rm AB}\sim 6.7$, $(1550-2500)\sim -0.38$,
$(FUV-NUV)_{\rm AB}\sim 0.42$ and $\beta_{\rm FUV}\sim -3.0$.

\subsubsection{Colour-colour diagrams}

\begin{figure*}
\epsfig{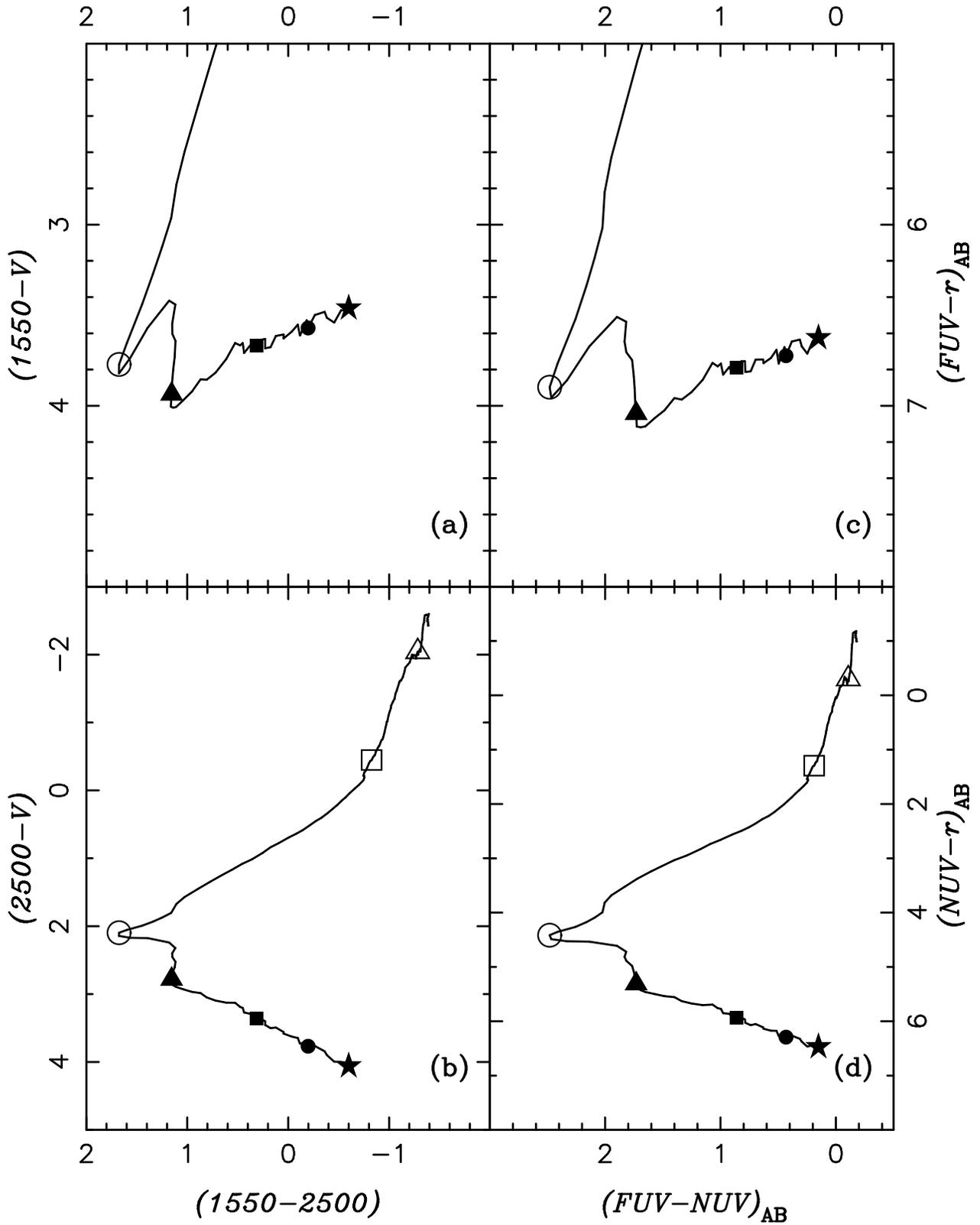}
\caption{
Colour-colour diagrams for the standard simulation set.
Ages are denoted by open triangles (0.01\,Gyr),
open squares (0.1\,Gyr), open circles (1\,Gyr),
filled triangles (2\,Gyr), filled squares (5\,Gyr),
filled circles (10\,Gyr) and filled stars (15\,Gyr).
}
\label{cc01}
\end{figure*}

\begin{figure*}
\epsfig{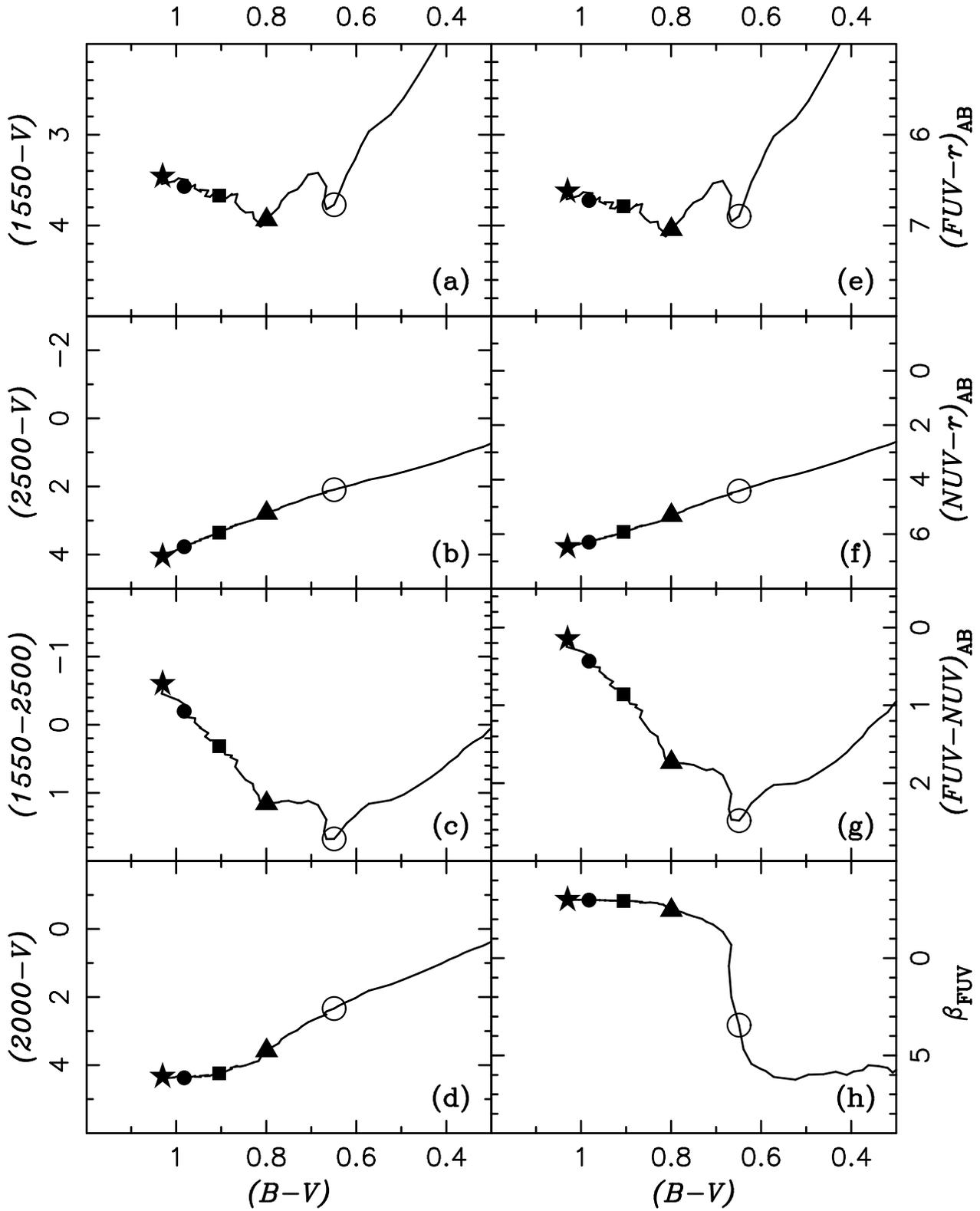}
\caption{
Similar to Figure~\ref{cc01}, but with $(B-V)$ as the abscissa.
}
\label{cc02}
\end{figure*}

Colour-colour diagrams are widely used as a diagnostic tool
in the study of stellar populations of early-type galaxies.
We present a few such diagrams in
Figures~\ref{cc01} and ~\ref{cc02} for the standard simulation set.
In these figures, most curves have a turning-point at
$\sim 1 \,{\rm Gyr}$, at which hot subdwarfs resulting
from binary interactions start to dominate the far-UV.

\subsubsection{The far-UV contribution for different
formation channels of hot subdwarfs}

\begin{figure*}
\epsfig{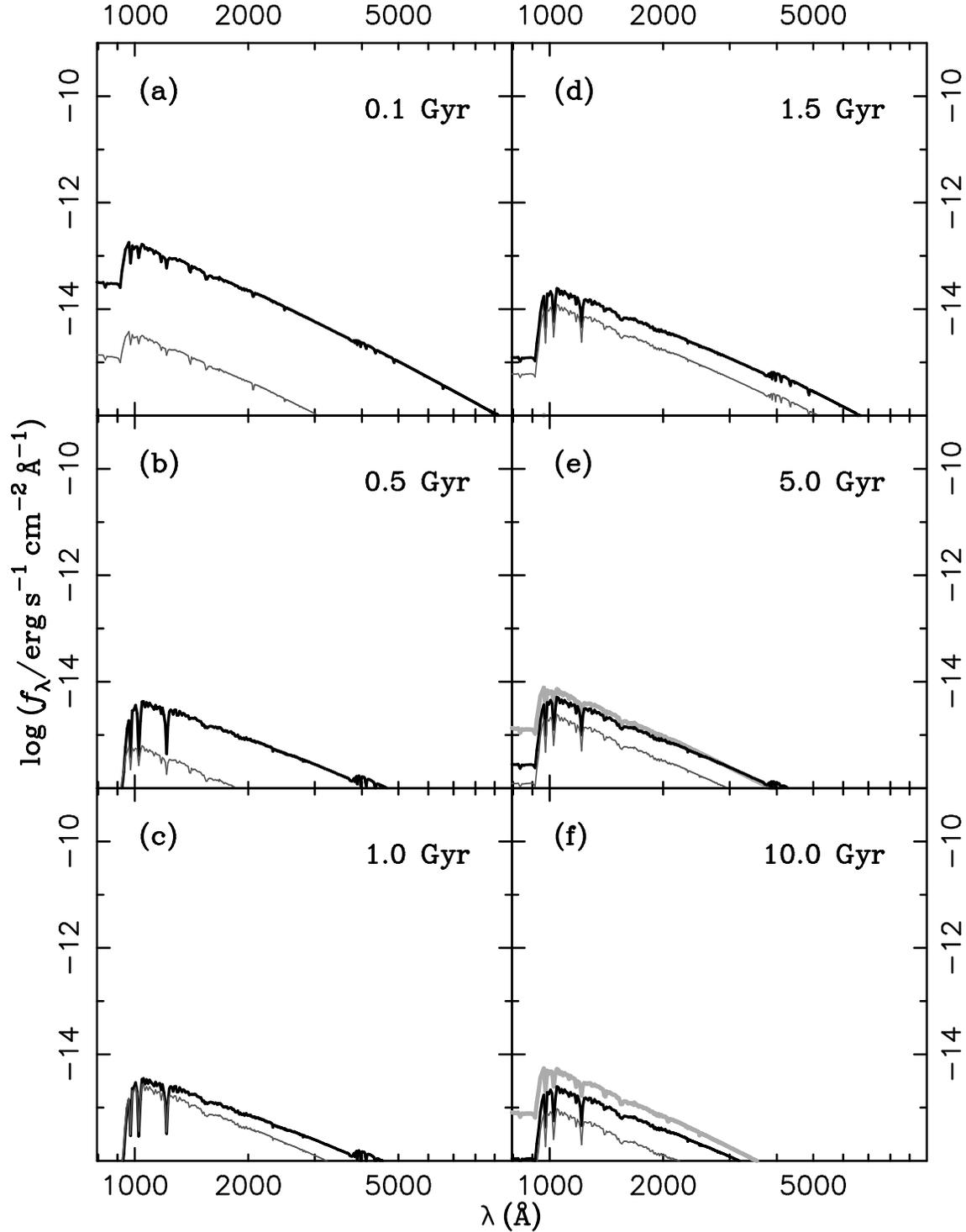}
\caption{ Integrated restframe intrinsic SEDs for hot subdwarfs for
different formation channels.  Solid, thin dark grey and thick 
light grey curves
are for the stable RLOF channel, the CE ejection channel and the
merger channel, respectively.  The merger channel starts to dominate
at an age of $t_{\rm SSP}\sim 3.5\,{\rm Gyr}$.  The figure is for the
standard simulation set with a stellar population mass of
$10^{10}M_\odot$ (including binaries), and the population is assumed
to be at a distance of 10\,Mpc.  }
\label{channel}
\end{figure*}

In our model, there are three channels for the formation of hot subdwarfs.
In the stable RLOF channel, the hydrogen-rich envelope of a star with a
helium core is removed by stable mass transfer, and helium is ignited
in the core. The hot subdwarfs from this channel are in
binaries with long orbital periods (typically $\sim 1000 {\,\rm d}$).
In the CE ejection channel, the envelope is ejected as a consequence
of the spiral-in in a CE phase. The resulting hot subdwarf binaries
have very short orbital periods (typically $\sim 1 {\,\rm d}$).
In the merger channel, a helium WD pair coalesces to produce a single
object. Hot subdwarfs from the merger channel are generally more massive
than those from stable RLOF channel or the CE channel and have
much thinner (or no) hydrogen envelope. They are therefore expected
to be hotter. See Han \etal \shortcite{han02,han03} for further details.

Figure~\ref{channel} shows the SEDs of the hot subdwarfs produced from
the different formation channels at various ages. It shows that hot
subdwarfs from the RLOF channel are always important, while the CE
channel becomes important at an age of $\sim 1 \,{\rm Gyr}$.  The
merger channel, however, catches up with the CE channel at $\sim 2.5
\,{\rm Gyr}$ and the stable RLOF channel at $\sim 3.5\,{\rm Gyr}$, and
dominates the far-UV flux afterwards.

\subsubsection{The effects of the model assumptions}

In order to systematically investigate the dependence of the UV-upturn
on the parameters of our model, we now vary the major model
parameters: $n(q')$ for the initial mass-ratio distribution, $q_{\rm
c}$ for the critical mass ratio for stable RLOF on the FGB or AGB,
$\alpha_{\rm CE}$ for the CE ejection efficiency and $\alpha_{\rm th}$
for the thermal contribution to the CE ejection. We carried out four
more simulation sets (Table~\ref{models}).  Figures~\ref{uvz1} and
~\ref{uvz2} show the UV-upturn evolution of the various simulation
sets. These figures show that the initial mass-ratio distribution is
very important.  As an extreme case, the mass-ratio distribution for
uncorrelated component masses (Set 2) makes the UV-upturn much weaker,
by $\sim 1 {\rm mag}$ in $(1550-V)$ or $(FUV-r)_{\rm AB}$, as compared
to the standard simulation set.  On the other hand, a rising
distribution (Set 3) makes the UV-upturn stronger.  Binaries with a
mass-ratio distribution of uncorrelated component masses tend to have
bigger values of $q$ (the ratio of the mass of the primary to the mass
of the secondary), and mass transfer is more likely to be unstable. As
a result, the numbers of hot subdwarfs from the stable RLOF channel
and the merger channel are greatly reduced, and the UV-upturn is
smaller in strength. Binaries with a rising mass-ratio distribution
tend to have smaller values of $q$ and therefore produce a larger
UV-upturn. A lower $q_{\rm c}$ (Set 4) leads to a weaker UV-upturn, as
the numbers of hot subdwarfs from the stable RLOF channel and the
merger channel are reduced. A higher CE ejection efficiency
$\alpha_{\rm CE}$ and a higher thermal contribution factor
$\alpha_{\rm th}$ (Set 5) result in an increase in the number of hot
subdwarfs from the CE channel, but a decrease in the number from the
merger channel, and the UV-upturn is not affected much as a
consequence.

\subsubsection{The importance of binary interactions}
\begin{figure*}
\epsfig{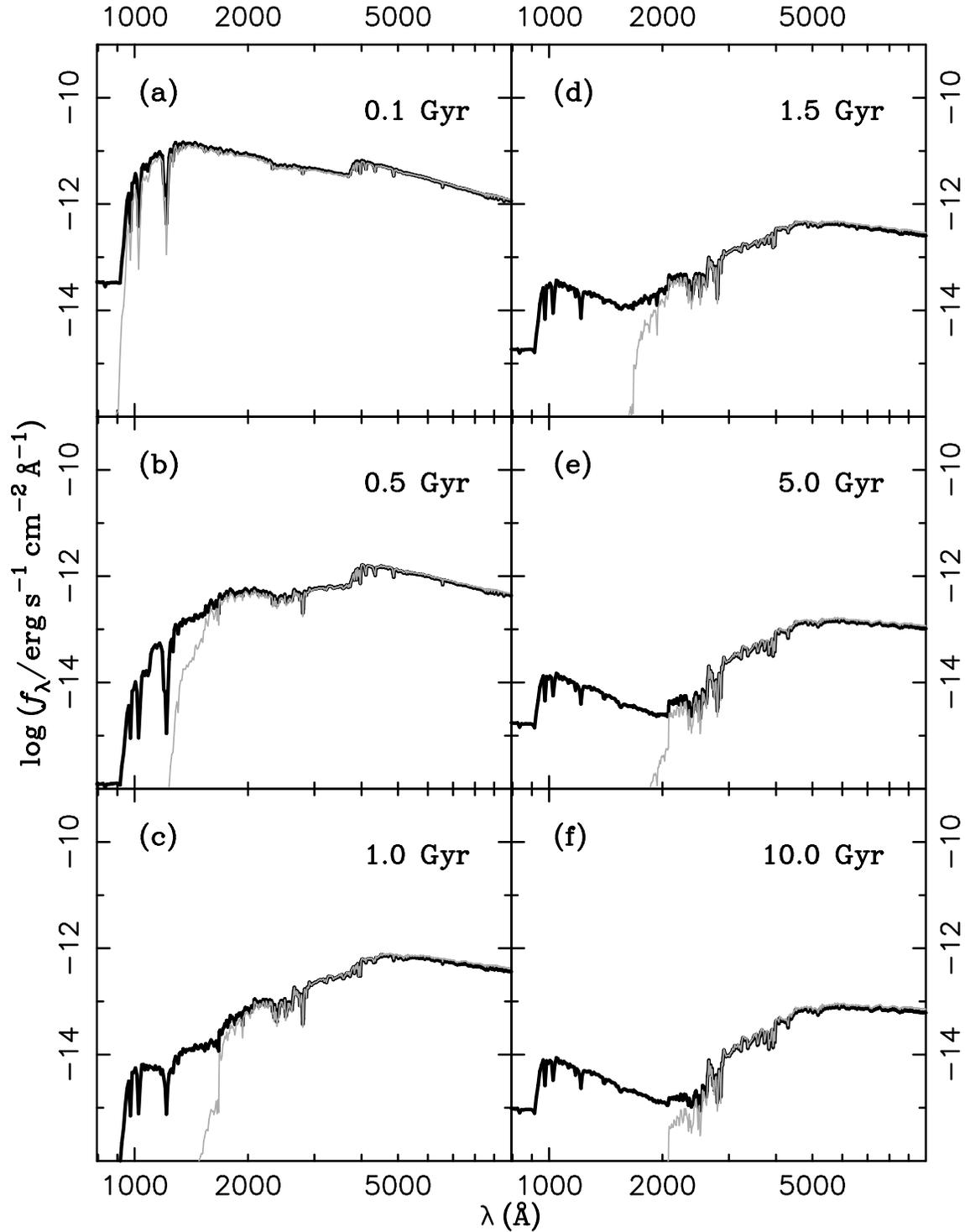}
\caption{
Integrated restframe intrinsic SEDs for a stellar population
(including binaries) with a mass of $10^{10}M_\odot$ at a distance of 
10\,Mpc.
Solid curves are for the standard simulation set
with binary interactions included,
and the light grey curves for the same population,
but no binary interactions are considered; the two components
evolve independently.
}
\label{binary}
\end{figure*}

\begin{figure*}
\epsfig{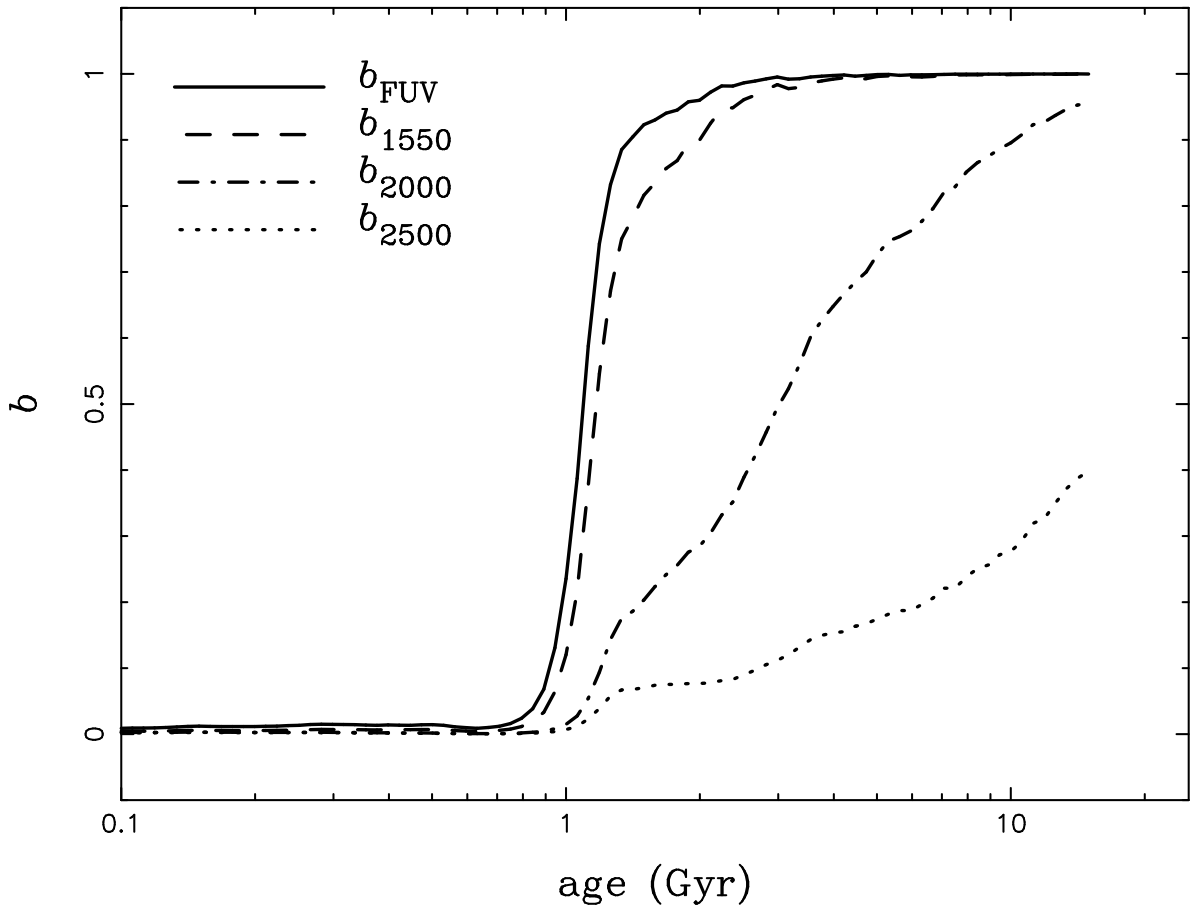}
\caption{
Time evolution of the fraction of the energy flux
in different UV wavebands originating from hot subdwarfs 
(and their descendants) formed in binaries
for the standard simulation set.
}
\label{bt}
\end{figure*}

Binaries evolve differently from single stars due to the occurrence of
mass transfer. Mass transfer may prevent mass donors from evolving to
higher luminosity and can produce hotter objects than expected in a
single-star population of a certain age (mainly hot subdwarfs and blue
stragglers\footnote{Blue stragglers are stars located on the main
sequence well beyond the turning-point in the colour-magnitude diagram
of globular clusters \cite{san53}, which should already have evolved
off the main sequence.  Collisions between low-mass stars and mass
transfer in close binaries are believed to be responsible for the
production of these hotter objects (e.g.\ Pols \& Marinus 1994, Chen
\& Han 2004, Hurley \etal 2005).}).  To demonstrate the importance of
binary interactions for the UV-upturn explicitly, we plotted SEDs of a
population for two cases in Figure~\ref{binary}.  Case 1 (solid
curves) is for our standard simulation set, which includes various
binary interactions, while case 2 (light grey curves) is for a
population of the same mass without any binary interactions.  The
figure shows that the hot subdwarfs produced by binary interactions
are the dominant contributors to the far-UV for a population older
than $\sim 1 \,{\rm Gyr}$. Note, however, that blue stragglers
resulting from binary interactions are important contributors to the
far-UV between 0.5 Gyr and 1.5 Gyr.

In order to assess the importance of binary interactions for the UV
upturn, we define factors that give the fraction of the flux in a
particular waveband that originates from hot subdwarfs 
produced in binaries: 
$b_{\rm FUV}=F^{\rm sd}_{\rm FUV}/F^{\rm total}_{\rm FUV}$, 
where $F^{\rm sd}_{\rm FUV}$ is the
integrated flux between 900\AA~ and 1800\AA~ radiated by hot subdwarfs
(and their descendants) produced by binary interactions, and
$F^{\rm total}_{\rm FUV}$ is the total integrated
flux between 900\AA~ and 1800\AA~.  We also defined other similar factors, 
$b_{\rm 1550}$, $b_{\rm 2000}$, and $b_{\rm 2500}$, for 
passbands of 1250\AA~ to 1850\AA~, 1921\AA~ to 2109\AA, and 2200\AA~
to 2800\AA, respectively.  Figure~\ref{bt} shows the time evolution of
those factors.  As the figure shows,
the hot-subdwarf contribution becomes increasingly important in the far-
and near-UV as the population ages.

\subsection{The model for composite stellar populations}

\begin{figure*}
\epsfig{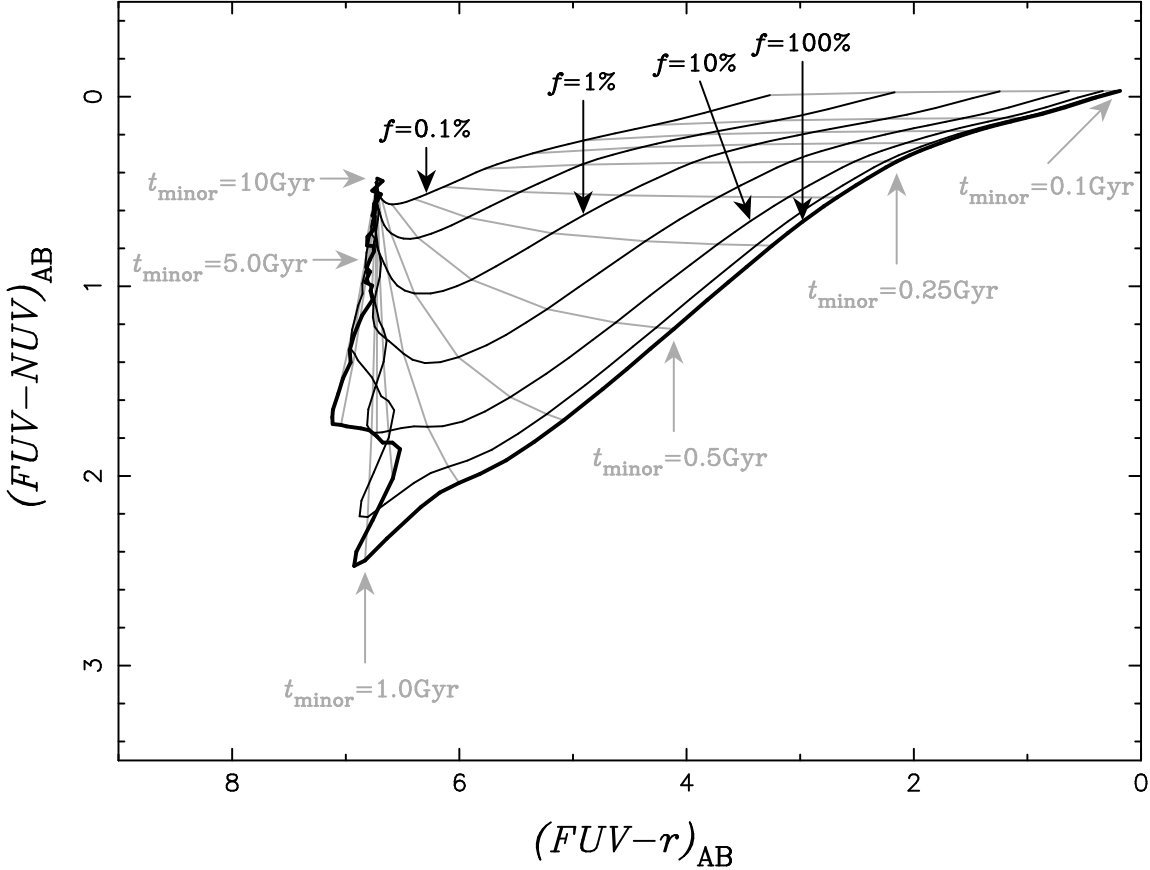}
\caption{
The diagram of $(FUV-NUV)_{\rm AB}$ versus $(FUV-r)_{\rm AB}$
for a composite stellar population (CSP) model
of elliptical galaxies with a major population
age of $t_{\rm major}=10\,{\rm Gyrs}$ (Set 6).
Solid curves are for given minor population fractions $f$
and are plotted in steps of $\Delta \log (f)=0.5$, as indicated.
Light grey curves are for fixed minor population ages $t_{\rm minor}$
and are plotted in steps of $\Delta \log (t_{\rm minor}/{\rm Gyr})=0.1$,
as indicated. The colours are presented in the restframe.
The thick solid curve for $f=100\%$ actually shows the evolution
of a simple stellar population with age $t_{\rm minor}$.
}
\label{frfn}
\end{figure*}

Early-type galaxies with a recent minor starburst can be modelled as a
composite stellar population (CSP). A CSP contains a major population
with an age of $t_{\rm major}$ and a minor population of age $t_{\rm
minor}$ and mass fraction $f$ (Section~3.3).  Figure~\ref{frfn} shows
the colour--colour diagram for CSPs with $t_{\rm major}=10 \,{\rm
Gyr}$ with varying $t_{\rm minor}$ and $f$. Note that the curves of
different $f$ start to converge to the SSP curve, the curve of
$f=100\%$, for $t_{\rm minor}>1 \,{\rm Gyr}$. This implies that there
exists a strong degeneracy between the age of the minor population and
the mass fraction.
%The figure could be used as a diagnostic tool in the
%study of early type galaxies with GALEX data.

\subsection{Theory versus observations}

\begin{figure*}
\epsfig{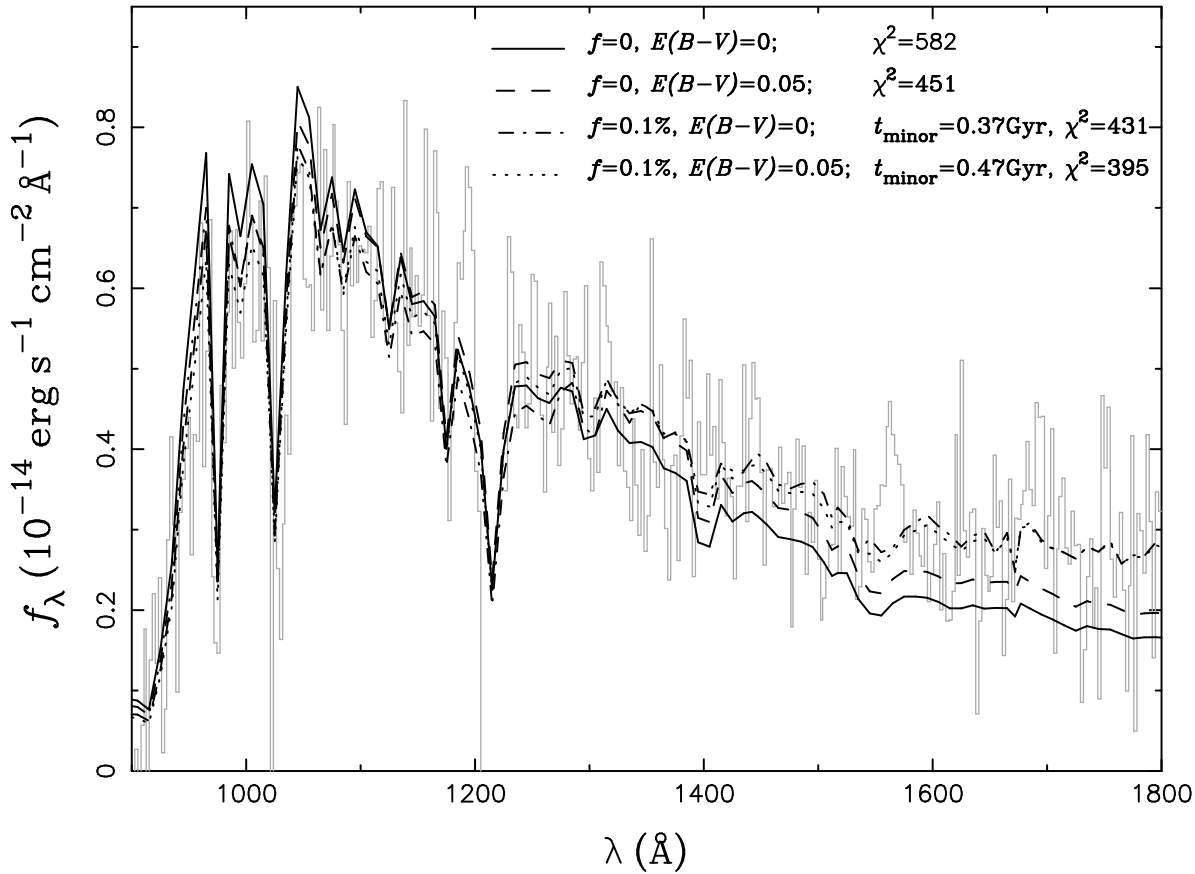}
\caption{
Far-UV SED fitting of NGC 3379, a standard
elliptical galaxy.  The grey histogram represents the HUT
observations of Brown \etal \shortcite{bro97} with 337 bins,
while the other curves are based on our
theoretical model with different assumptions about a minor recent
population and different amounts of extinction. The age of the old,
dominant population, $t_{\rm major}$, is
assumed to be 10\,Gyr in all cases. The minor
population, making up a total fraction $f$ of the stellar mass of the
galaxy, is assumed to have experienced a starburst
$t_{\rm minor}$ ago, as indicated,
lasting for 0.1\,Gyr. To model the effects of dust
extinction, we applied the internal dust extinction model of Calzetti
\etal \shortcite{cal00}.
For a given $f$ and $E(B-V)$, we varied the galaxy mass and the age of
the minor population.  For each case, the curves give the fits
obtained. Note that the best fits require the presence of a minor young
population.
}
\label{fitting}
\end{figure*}

\begin{figure*}
\epsfig{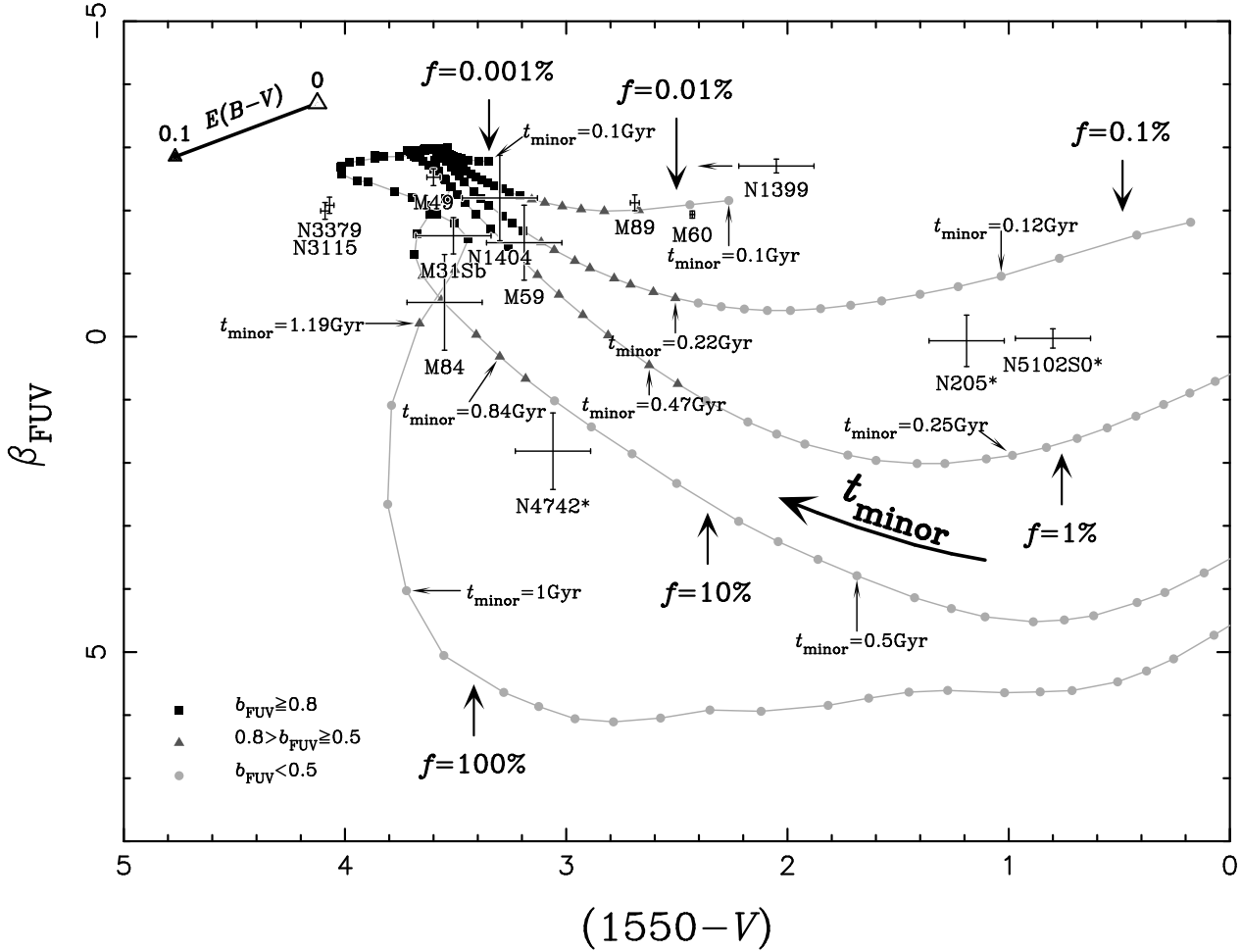}
\caption{
Evolution of far-UV
properties [the slope of the far-UV spectrum, $\beta_{\rm FUV}$,
versus $(1550-V)$] for a composite stellar population (CSP) model of
elliptical galaxies with a major population
age of $t_{\rm major}=10\,{\rm Gyrs}$ (Set 6).
The mass fraction of the younger
population is denoted as $f$ and the time since the formation as
$t_{\rm minor}$ [squares, triangles or dots are 
plotted in steps of $\Delta \log (t)=0.025$].
Note that the model for $f=100\%$ 
shows the evolution of a simple stellar population with age
$t_{\rm minor}$.  The legend is for $b_{\rm FUV}$, which is
the fraction of the UV flux that originates from hot subdwarfs resulting
from binary interactions. The effect of internal extinction is
indicated in the top-left corner, based on the Calzetti internal
extinction model with $E(B-V)=0.1$ (Calzetti \etal, 2000).
For comparison, we also plot galaxies with error bars from HUT
(Brown \etal, 1997) and IUE observations
(BBBFL).  The galaxies with strong signs of
recent star formation are denoted with an asterisk (NGC 205, NGC 4742,
NGC 5102).
}
\label{beta}
\end{figure*}

\begin{figure*}
\epsfig{file=15v.ps,angle=270,width=17cm}
\caption{
The diagrams of $(1550-V)$ versus $(1550-2500)$ (a) and
$(2000-V)$ versus $(1550-2500)$ (b)
for a composite stellar population (CSP) model of elliptical galaxies
with a major population
age of $t_{\rm major}=10\,{\rm Gyrs}$ (Set 6).
Solid curves are for given minor population fractions $f$
and are plotted, from left to right,
in steps of $\Delta \log (f)=0.5$, as indicated.
Light grey curves are for fixed minor population ages $t_{\rm minor}$
and are plotted, from top to bottom,
in steps of $\Delta \log (t_{\rm minor}/{\rm Gyr})=0.1$ , as indicated.
Note that the colours are given in the restframe and intrinsic.
The thick solid curve for $f=100\%$ actually shows the evolution
of a simple stellar population with age $t_{\rm minor}$.
Solid squares are for quiescent early-type galaxies observed
with IUE by BBBFL and the data points are
taken from Dorman, O'Connell \& Wood \shortcite{dor95}.
}
\label{15v}
\end{figure*}

\begin{figure*}
\epsfig{file=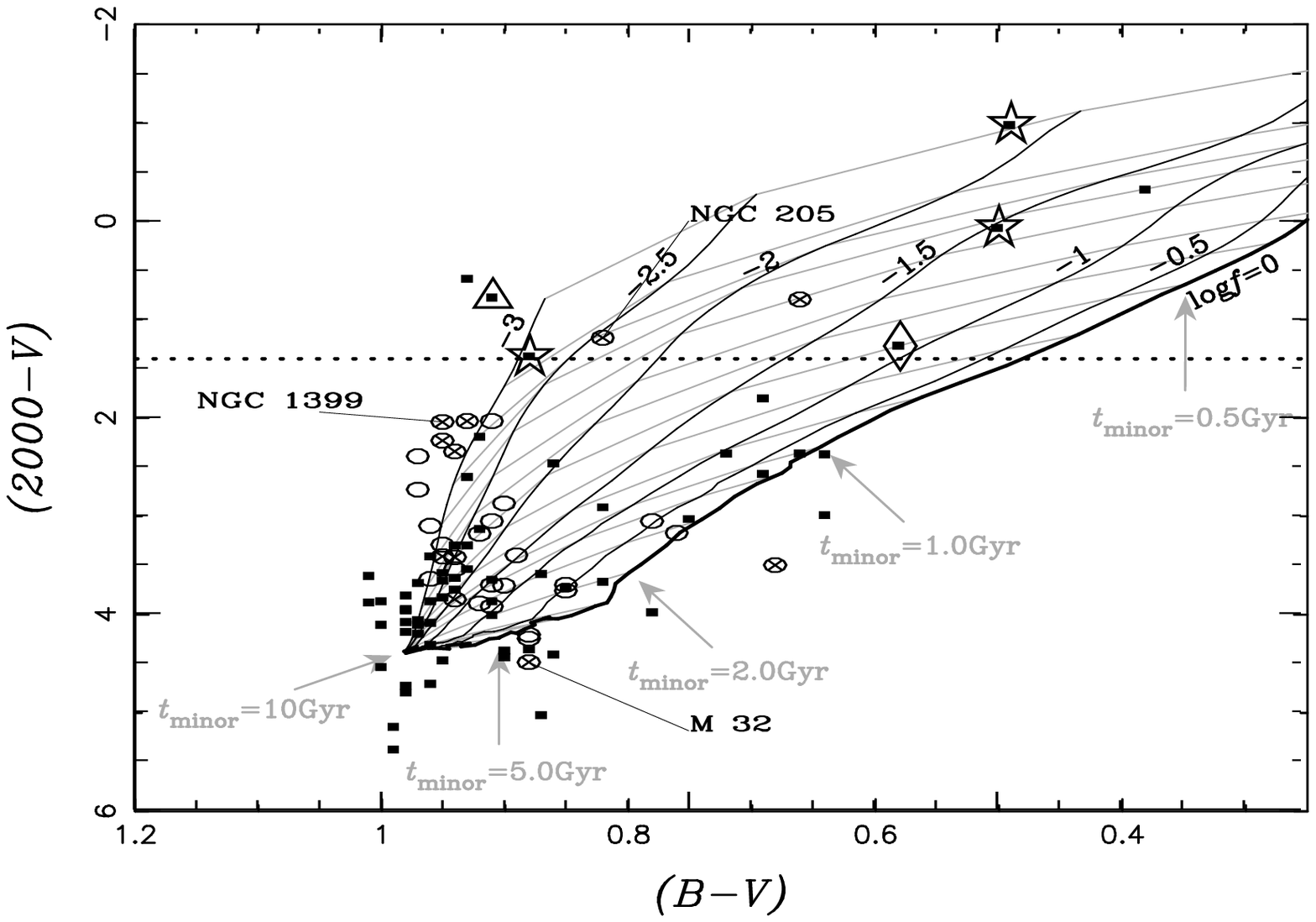,width=17cm}
\caption{
The diagrams of $(B-V)$ versus $(2000-V)$
for a composite stellar population (CSP) model of elliptical galaxies
with a major population
age of $t_{\rm major}=10\,{\rm Gyrs}$ (Set 6).
Solid curves are for given minor population fractions $f$
and are plotted, from left to right,
in steps of $\Delta \log (f)=0.5$, as indicated.
Light grey curves are for fixed minor population ages $t_{\rm minor}$
and are plotted, from top to bottom,
in steps of $\Delta \log (t_{\rm minor}/{\rm Gyr})=0.1$, as indicated.
The thick solid curve for $f=100\%$ actually shows the evolution
of a simple stellar population with age $t_{\rm minor}$.
Note that the colours are intrinsic and are plotted in the restframe.
Overlayed on this diagram is figure~2 of Deharveng, Boselli \& Donas
\shortcite{deh02}, in which solid squares are for their sample
observed with the FOCA experiment and open circles (including
circles with crosses) are for the sample of BBBFL observed
with IUE. Crosses denote objects that have been studied in
detail with HUT or HST. For galaxies bluer than $(2000-V)=1.4$,
solid squares with big stars and circles with crosses are
for galaxies with recent star formation.
NGC 4168 (solid square with a big triangle) has a low-luminosity Seyfert
nucleus. CGCG 119030 (solid square with a big diamond) could be
misclassified as an elliptical, as it 
is classified as a spiral in the NASA/IPAC Extragalactic Database (NED).
}
\label{20v}
\end{figure*}

\begin{figure*}
\epsfig{file=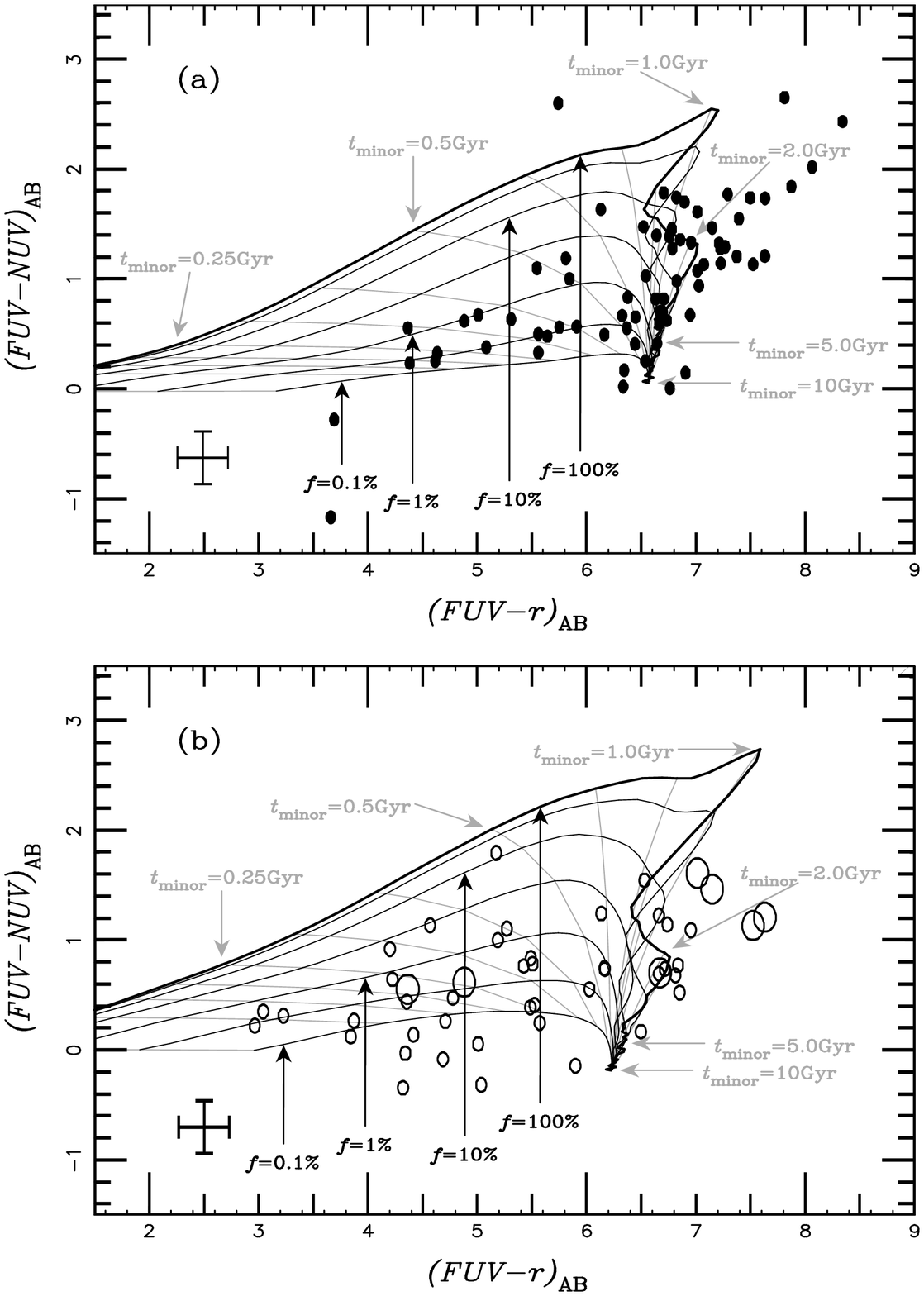,width=14cm}
\caption{
The diagrams of $(FUV-r)_{\rm AB}$ versus $(FUV-NUV)_{\rm AB}$
for a composite stellar population (CSP) with a major population
age of $t_{\rm major}=10\,{\rm Gyrs}$ (Set 6).
Solid curves are for given minor population fractions $f$
and are plotted, from bottom to top,
in steps of $\Delta \log (f)=0.5$, as indicated.
Light grey curves are for fixed minor population ages $t_{\rm minor}$
and are plotted, from left to right,
in steps of $\Delta \log (t_{\rm minor}/{\rm Gyr})=0.1$, as indicated.
The thick solid curve for $f=100\%$ actually shows the evolution
of a simple stellar population with age $t_{\rm minor}$.
Note that the colours are intrinsic but not in the restframe.
Panel (a) assumes a redshift of $z=0.05$ and panel (b)  $z=0.15$.
Overlayed on this diagram is figure~3 of Rich \etal \shortcite{ric05},
in which circles are for their sample
observed with GALEX for quiescent early-type galaxies.
Filled circles (panel (a)) are observational data points for galaxies of
$0<z<0.1$ and open circles (panel (b)) for $0.1<z<0.2$.
}
\label{fr}
\end{figure*}

\subsubsection{The fitting of the far-UV SED}

In our binary population synthesis model, we adopted solar metallicity.
To test the model, we chose NGC 3379, a typical elliptical galaxy with
a metallicity close to solar \cite{gre04} to fit the far-UV SED.
Figure~\ref{fitting} presents various fits
that illustrate the effects of different
sub-populations with different ages and different amounts of assumed
extinction. As the figure shows, acceptable fits can be obtained for
the various cases.
Our best fits require the presence of a sub-population of
relatively young stars with an age $< 0.5\,$Gyr,
making up $\sim 0.1$\,\% of the 
total population mass.

The existence of a relatively young population could imply the
existence of some core-collapse supernovae in these galaxies. 
If we assume that an elliptical 
galaxy has a stellar mass of $10^{11}M_\odot$ and $0.1\%$ of the mass
was formed {\it during} the last 0.4\,Gyr, then a mean star formation
rate would be $0.25M_\odot/{\rm yr}$, about one tenth 
that of the Galaxy.  Core-collapse supernovae would be possible. Indeed,
a Type Ib supernova, SN 2000ds, was discovered in NGC 2768, an elliptical
galaxy of type E6 \cite{van03}.
 
\subsubsection{The UV-upturn magnitudes versus the
far-UV spectral index}

There is increasing evidence that many elliptical galaxies had some
recent minor star-formation events \cite{sch06,kav06}, which also
contribute to the far-UV excess.  To model such secondary minor
starbursts, we have constructed CSP galaxy models,
consisting of one old, dominant population with an assumed age 
$t_{\rm major}=10\,{\rm Gyr}$ and a younger population of variable age, 
making up a fraction $f$ of the stellar mass of the system.
Our spectral modelling shows that a recent minor
starburst mostly affects the slope in the far-UV SED,
and we therefore defined a far-UV slope index $\beta_{\rm FUV}$ 
(Section~2.4). In order to assess the importance of binary interactions, 
we also defined a binary contribution factor 
$b_{\rm FUV}$ (Section~4.1.6), which is the fraction of far-UV
flux radiated by hot subdwarfs produced by binary interactions.

Figure~\ref{beta} shows the far-UV slope as a function of UV excess, a
potentially powerful diagnostic diagram which illustrates how the UV
properties of elliptical galaxies evolve with time in a dominant old
population with a young minor sub-population.  For comparison, we also
plot observed elliptical galaxies.
Some of the observed galaxies are from Astro-2 observations with
an aperture of $10'' \times 56''$ \cite{bro97}, some
are from IUE observations with an aperture of $10'' \times 20''$ 
\cite{bur88}.
The value of $\beta_{\rm FUV}$ for NGC 1399, however, is derived from
Astro-1 HUT observation with an aperture of $9''.4 \times 116''$
\cite{fer91}, and its $(1550-V)$ comes from the IUE observations.
As the far-UV light is more concentrated toward the centre of the
galaxy than the optical light \cite{ohl98}, the value of $(1550-V)$ for
NGC 1399 should be considered an upper limit for the galaxy area
covered by the observation. The galaxies plotted are all
elliptical galaxies except for NGC 5102, which is 
a S0 galaxy, and the nucleus of M~31, which is a Sb 
galaxy.  Active galaxies or
galaxies with large errors in $\beta_{\rm FUV}$ from BBBFL
have not been plotted.

Overall, the model covers the observed range of properties reasonably
well.  Note in particular that the majority of galaxies lie in the
part of the diagram where the UV contribution from binaries is
expected to dominate (i.e. where $b_{\rm FUV}> 0.5$).  The location of
M~60 and M~89 in this figure implies $f\sim 0.01\%$ and $t_{\rm
minor}\sim 0.11\, {\rm Gyr}$ with $b_{\rm FUV}\sim 0.5$.
Interestingly, inspection of the HUT spectrum of M~60 (see the mid-left
panel of figure~3 in Brown \etal \shortcite{bro97}) shows the presence
of a marginal C~IV absorption line near 1550\AA.  Chandra observations
show that M~89 has a low luminosity AGN \cite{xu05}. This would
make $(1550-V)$ bluer and may also provide indirect evidence for low levels
of star formation.

The galaxy NGC 1399 requires special mention, as it is UV-bright and
the young star-hypothesis was believed to have been ruled out due to
the lack of strong C~IV absorption lines in its HUT spectrum
\cite{fer91}.  However, any young star signature, if it exists, would
have been diluted greatly in the HUT spectrum, as the aperture
of the HUT observation is much larger than that of the IUE
observation covering mainly the galaxy nucleus.

Our model is sensitive to both low levels and high levels of star
formation.  It suggests that elliptical galaxies
had some star formation activity in the relatively recent past
($\sim 1\,$Gyr ago).  AGN and supernova activity may provide
qualitative supporting evidence for this picture, 
since the former often appears
to be accompanied by active star formation, while supernovae, both
core collapse and thermonuclear, tend to occur mainly within
1\,--\,2\,Gyr after a starburst in the most favoured supernova models.
In Figure~\ref{beta}, we have plotted 13 early-type galaxies
altogether. Using the Padova-Asiago supernova online
catalogue\footnote{{\it http://web.pd.astro.it/supern/snean.txt}}, which
lists supernovae recorded ever since 1885, we found 8 supernovae
in six of the galaxies: SN 1885A (Type I) in M~31, 
SN 1969Q (type unavailable) in M~49, 
SN 2004W (Type Ia) in M~60,
SN 1939B (Type I) in M~59, SN 1957B (Type Ia), SN 1980I (Type Ia),
SN 1991bg (Type Ia) in M~84 and
SN 1935B (type unavailable) in NGC 3115. 
The majority of supernovae in these galaxies appear to be of Type Ia.

\subsubsection{Colour-colour diagrams}

Similar to the above subsection, we have constructed composite stellar
population models for elliptical galaxies consisting of a 
major old population ($t_{\rm major}=10\,{\rm Gyr}$) 
and a minor younger population, but for colour-colour diagrams.

Figure~\ref{15v} shows the appearance of a galaxy in the colour-colour
diagrams of $(1550-V)$ versus $(1550-2500)$ (panel (a))
and $(2000-V)$ versus $(1550-2500)$ (panel (b))
in the CSP model for different fractions $f$ and
different ages of the minor population.
Solid squares in panel (a) of the figure
are for quiescent early-type galaxies observed
with IUE by BBBFL and the data points are
taken from table~2 of Dorman, O'Connell \& Wood \shortcite{dor95}.
The observed data points are located in a region
without recent star formation or with a very low level of
recent star formation ($f\la 0.1\%$). Therefore, the
observations are naturally explained with our model.
Panel (a) shows the epoch when
the effects of the starburst fade away, leading to a fast evolution of
the galaxy colours, and the diagram therefore provides a potentially 
powerful
diagnostic to identify a minor starburst in an otherwise old
elliptical galaxy that occurred up to $\sim 2\,$Gyr ago.  For larger
ages, the curves tend to converge in the 
$(1550-2500)$ versus $(1550-V)$
diagram. 
Note that this is not so much the case in the $(1550-2500)$ versus
$(2000-V)$ diagram, which therefore could provide better diagnostics.

Figure~\ref{20v} is a diagram of $(B-V)$ versus $(2000-V)$ for
a CSP with a major population age $t_{\rm major}=10 \,{\rm Gyr}$ and
variable minor population age $t_{\rm minor}$ and
various minor population mass fractions $f$. Overlayed on this
diagram is figure~2 of Deharveng, Boselli \& Donas \shortcite{deh02}
for observational data points of early-type galaxies.
NGC 205 and NGC 5102
(the circles with crosses above the line $(2000-V)=1.4$)
are known to have direct evidence of massive star formation
\cite{hod73,pri79};
therefore Deharveng, Boselli \& Donas \shortcite{deh02} individually
examined the seven galaxies with $(2000-V)<1.4$ in their
sample for suspected star formation.
CGCG 119053, CGCG 97125, VCC 1499 (the three solid squares with big
stars) showed hints of star formation,
NGC 4168 (the solid square with a big triangle)  has a low-luminosity
Seyfert nucleus and CGCG 119030 (the solid square with a big diamond)
could be a spiral galaxy instead of an elliptical galaxy.
However, no hint of star formation has been found for
VCC 616 (the solid square on the far-left above the 
line $(2000-V)=1.4$) and CGCG 119086 (the solid square on the far-right
above the line $(2000-V)=1.4$).
Our model can explain the observations satisfactorily except for
CGCG 119086, which needs further study.

Figure~\ref{fr} shows the diagrams of $(FUV-r)_{\rm AB}$ versus
$(FUV-NUV)_{\rm AB}$ for a CSP galaxy model with a major population
age $t_{\rm major}=10 \,{\rm Gyr}$ and variable minor population age
$t_{\rm minor}$ and various minor population mass fractions $f$.  In
these diagrams, the colours are not shown in the restframe, but have
been redshifted (i.e. the wavelength is $(1+z)$ times the restframe
wavelength, where $z$ is the redshift); panel (a) is for a redshift of
$z=0.05$ and panel (b) for $z=0.15$.  Overlayed on the two panels are
quiescent early-type galaxies observed with GALEX by Rich \etal
\shortcite{ric05}.  The observed galaxies are for a redshift range
$0<z<0.1$ (panel (a)) and $0.1<z<0.2$ (panel (b)). We note that most
of the quiescent galaxies are located in the region with $f\la 1\%$.

In Figures ~\ref{15v} to ~\ref{fr}, we adopted a major population age
of $t_{\rm major}=10\, {\rm Gyr}$, and the colours are intrinsic.
However, adopting a different age for the major population can change
the diagrams; for example, a larger age leads to bluer $(1550-2500)$ or
$(FUV-NUV)$ colours.  In contrast, internal dust extinction 
shifts the curves towards redder colours \cite{cal00}.
Considering the uncertainties in the modelling, we take our model to
be in reasonable agreement with the observations.

\subsubsection{UV-upturn magnitudes and their evolution with redshift}

There is an observed spread in the $(1550-V)$, $(1550-2500)$ and
$(2000-V)$ colours of early-type galaxies.  As can be seen from
Figures~\ref{beta}, ~\ref{15v}, 
and ~\ref{20v}, this spread
is satisfactorily explained by our model.

Brown \etal \shortcite{bro03} showed with HST observations that the
UV-upturn does not evolve much with redshift, a result apparently
confirmed by Rich \etal \shortcite{ric05} with GALEX observation of a
large sample.  This is contrary to the prediction of both the
metal-poor and the metal-rich model, as both models require a large age for
the hot subdwarfs and therefore predict that the UV-upturn should decline
rapidly with redshift.  Our binary model, however, predicts that
that UV-upturn does not evolve much with redshift (see
Figures~\ref{uvz1} and ~\ref{uvz2}), consistent with the recent
observations.

Lee \etal \shortcite{lee05} and Ree \etal \shortcite{ree07} 
studied the look-back time evolution of the UV-upturn from 
the brightest elliptical galaxies in 12 clusters at redshift $z<0.2$
with GALEX. Compared to local giant elliptical galaxies, 
they found that the UV-upturn of the 12 galaxies is redder. 
However, the local giant elliptical galaxies are quite special.
NGC 1399 and M~87, with the strongest UV-upturn,
have the largest known specific frequencies
of globular clusters \cite{ost93}, and M~87 hosts an active galactic
nuclei (AGN) with the best-known jet \cite{cur18,wat05}. 
Given a larger sample of elliptical galaxies, no matter how luminous 
they are, and a bigger redshift range, the UV-upturn 
is not found to
decline with redshift. 

\subsubsection{Implication for star formation history of early-type galaxies}

Boselli \etal \shortcite{bos05} studied the UV properties of 264 early-type
galaxies in the Virgo cluster with GALEX. They showed that 
$(FUV-NUV)_{\rm AB}$ ranges from 3 to 0, consistent with the theoretical
range shown in panel (b) of Figure~\ref{uvz2}.
The colour index $(FUV-NUV)_{\rm AB}$ of those galaxies
becomes bluer with luminosity from 
dwarfs ($L_{\rm H}\sim 10^8L_{{\rm H,}\odot}$) to 
giants ($L_{\rm H}\sim 10^{11.5}L_{{\rm H,}\odot}$),
i.e. a luminous galaxy tends to have a bluer $(FUV-NUV)_{\rm AB}$.
Panel (b) of Figure~\ref{uvz2} shows that
$(FUV-NUV)_{\rm AB}$ becomes bluer with population age 
for $t_{\rm SSP}> 1\,{\rm Gyr}$. 
Taking the stellar populations as an ``averaged'' SSP,
we may conclude that a luminous early-type galaxy is older, 
or in other words, the less luminous an early-type galaxy is,
the younger the stellar population is or the later the population
formed.

\subsection{The UV-upturn and metallicity}

As far as we know, metallicity does
not play a significant role in the mass-transfer process or the
envelope ejection process for the formation of hot subdwarfs,
although it may affect the properties of the binary population
in more subtle ways. We therefore expect that 
$(FUV-r)_{\rm AB}$ or $(1550-V)$ from our model is not very 
sensitive to the metallicity of the population,
This is in agreement with the recent large sample
of GALEX observations by Rich \etal \shortcite{ric05}. 
Boselli \etal \shortcite{bos05} 
and Donas \etal \shortcite{don06} studied nearby early-type
galaxies with GALEX. Neither of them show that
$(FUV-r)_{\rm AB}$ correlates significantly with metallicity,  
However, both of them found a positive correlation 
between $(FUV-NUV)_{\rm AB}$ and metallicity. 
This can possibly be explained with our model. 
The UV-upturn magnitudes $(FUV-r)_{\rm AB}$ or $(1550-V)$ does
not evolve much with age for $t_{\rm SSP}>1\,{\rm Gyr}$
while $(FUV-NUV)_{\rm AB}$ decreases significantly with age
(see Figures~\ref{uvz1} and ~\ref{uvz2}).
A galaxy of high metallicity may have a larger age and 
therefore a stronger $(FUV-NUV)_{\rm AB}$.

\subsection{Comparison with previous models}

Both metal-poor and metal-rich models are quite {\it ad hoc} and
require a large age for the hot subdwarf population (Section~1),
which implies that the UV-upturn declines rapidly with look-back time
or redshift. In our model, hot subdwarfs are produced naturally by
envelope loss through binary interactions, which do not depend
much on the age of the population older than $\sim 1 \,{\rm Gyr}$, and
therefore our model predicts little if any evolution of the
UV-upturn with redshift. Note, however, that $(FUV-NUV)_{\rm AB}$
declines significantly more with redshift than $(FUV-r)_{\rm AB}$, as
the contribution to the near-UV from blue stragglers resulting from
binary interactions becomes less important for an older population.

The metal-rich model predicts a positive correlation between the
magnitude of the UV-upturn and metallicity; for example,
$(1550-V)$ correlates
with metallicity.  However, such a correlation is not expected from
our binary model as metallicity does not play an essential role in the
binary interactions. Furthermore, even though the metal-rich model
could in principle account for the UV-upturn in old, metal-rich giant
ellipticals, it cannot produce a UV-upturn in lower-metallicity dwarf
ellipticals. In contrast, in a binary model, the UV-upturn is
universal and can account for UV-upturns from dwarf ellipticals to
giant ellipticals.

\section{Summary and Conclusion}

By applying the binary scenario of Han \etal \shortcite{han02,han03}
for the formation of hot subdwarfs, we have developed
an evolutionary population synthesis model for the
UV-upturn of elliptical galaxies based on a first-principle approach.
The model is still quite simple and does not
take into account more complex star-formation histories, possible
contributions to the UV from AGN activity, non-solar metallicity or
a range of metallicities.  Moreover, the binary population
synthesis is sensitive to uncertainties in the binary
modelling itself, in particular the mass-ratio distribution and the
condition for stable and unstable mass transfer \cite{han03}. We have
varied these parameters and found these uncertainties do not
change the qualitative picture, but affect some of the quantitative
estimates.

Despite its simplicity, our model can successfully reproduce most of
the properties of elliptical galaxies with a UV excess: the
range of observed UV excesses, both in $(1550-V)$ and $(2000-V)$
(e.g.\ Deharveng, Boselli \& Donas, 2002),
and their evolution with redshift.  The model predicts
that the UV excess is not a strong function of age, and hence is not a
good indicator for the age of the dominant old population, as has been
argued previously \cite{yi99}, but is very consistent with recent
GALEX findings \cite{ric05}.  We typically find that the $(1550-V)$
colour changes rapidly over the first 1\,Gyr and only varies slowly
thereafter. This also implies that all old galaxies should show a UV
excess at some level. Moreover, we expect that the model is not very
sensitive to the metallicity of the population. 
The UV-upturn is therefore expected to be universal.

Our model is sensitive to both low levels and high levels of star
formation.  It suggests that elliptical galaxies
had some star formation activity in the relatively recent past
($\sim 1\,$Gyr ago).  AGN and supernova activity may provide
supporting evidence for this picture.

The modelling of the UV excess presented in this study is only a
starting point: with refinements in the spectral modelling, including
metallicity effects, and more detailed modelling of the global
evolution of the stellar population in elliptical galaxies, we propose
that this becomes a powerful new tool helping to unravel the
complex histories of elliptical galaxies that a long time ago looked
so simple and straightforward.

\section*{Acknowledgements}
We are grateful to an anonymous referee for 
valuable comments
which help to improve the presentation, to
Kevin Schawinski for numerous discussions and
suggestions, to Thorsten Lisker for insightful comments
leading to Section 4.3.5.
This work was in part supported by the Natural Science
Foundation of China under Grant Nos.\ 10433030 and 10521001, the
Chinese Academy of Sciences under Grant No.\ KJCX2-SW-T06 (Z.H.), and
a Royal Society UK-China Joint Project Grant (Ph.P and Z.H.).

\end{document}